\documentclass[twocolumn,preprintnumbers,amsmath,amssymb]{revtex4}
\usepackage{amsmath}
\usepackage{graphicx}% Include figure files
\usepackage{dcolumn}% Align table columns on decimal point
\usepackage{bm}
\usepackage{color}
\input cyracc.def

\begin{document}

\title{How to Measure Significance of Community Structure in Complex Networks}

\author{Yanqing Hu$^1$\footnote{yanqing.hu.sc@gmail.com}, Yiming Ding$^2$, Ying Fan$^1$, Zengru Di$^1$\footnote{zdi@bnu.edu.cn}}
 \affiliation{1. Department of Systems Science, School of Management and Center for Complexity
 Research, Beijing Normal University, Beijing 100875, China
\\2. Wuhan Institute of Physics and Mathematics,
The Chinese Academy of Sciences, Wuhan 430071, China
 }

\begin{abstract}
\end{abstract}


 %check!!!

\maketitle \textbf{Community structure analysis is a powerful tool
for complex networks, which can simplify their functional analysis
considerably. Recently, many approaches were proposed to community
structure detection, but few works were focused on the significance
of community structure. Since real networks obtained from complex
systems always contain error links, and most of the community
detection algorithms  have random factors,  evaluate the
significance of community structure is important and urgent. In this
paper, we use the eigenvectors' stability to characterize the
significance of community structures. By employing the eigenvalues
of Laplacian matrix of a given network, we can evaluate the
significance of its community structure and obtain the optimal
number of communities, which are always hard for community detection
algorithms. We apply our method to many real networks. We find that
significant community structures exist in many social networks and
C.elegans neural network, and that less significant community
structures appear in protein-interaction networks and metabolic
networks. Our method can be applied to broad clustering problems in
data mining due to its solid mathematical basis and efficiency.}

Complex networks have become a general tool for the analysis of
complex systems with many interacting elements. The study of the
community structure is of great importance for complex networks (see
\cite{SantoArxiv} as a review). Commonly in many real-world
networks, some small subnetworks (communities) have more connections
within themselves; but comparatively, they are less likely to be
connected with the rest parts. Since nodes in a tight-knit
subnetwork have more properties in common, divide the network into
such communities could simplify the functional analysis
considerably. As a result, the identification of community structure
has been the focus of many recent efforts. Generally speaking, such
an identification  contains two problems: One is  to detect the
community structure, which was extensively studied during the recent
5 years
\cite{SantoArxiv,linear_time,newman_spectra,GN,Systematic,WLC,Newman_Q}.
The second is  to evaluate its (community structure) significance,
which was hardly settled by researchers in the past. We believe that
some networks have clear communities while others don't. But whether
the community structure exists in the network or not, almost all
algorithms could find its ``community structure"; many algorithms
can even find  community structures in random networks, which are
essentially nonexistent at all.  Besides, many real-world networks
contain some error links and algorithms of detecting community
structure have some random factors \cite{Fan}. How to evaluate the
effects of error links and random factors in the community
structure? Therefore, the evaluation of the significance of
community structure  is imperative. Given a network, it is
meaningless to detect the community when the community structure is
not significate or when just few error links can considerably change
the community structure detected.

In previous works, only a few methods \cite{Newman_Q,Gfeller,Yhu}
can evaluate the significance of community structure, and all of
them require to know the community structure  before the evaluation.
However, the significance of community structure should be the
property of network itself, which is independent of the partition
algorithm, and can be evaluated without knowing the exact
communities. According to the well studied bi-communities of network
\cite{Newman_Q}, to calculate the significance of community
structure can be transformed to measure the stability of
eigenvectors. In the following sections, we will extend the
bi-communities problem to multi-communities problem and design an
index to evaluate the significance of the community structure.
Furthermore, we apply the method to many types of networks. We find
that C. elegans neural network and social networks usually have
distinct community structure, while metabolic networks and
protein-interaction networks don't. The results are consistent with
our previous research \cite{Yhu}.

\section{Method}
How to evaluate the impact of error links and random factors of
algorithm? The two aspects can be merged into one problem. We can
regard the random factors of algorithms as error link liked cases.
That is, we can suggest that all random factors are caused by error
links. If the community structure is very clear, a few error links
will not impact the structure greatly, neither will the random
factors of algorithm \cite{Fan}. Otherwise, if the community
structure is fuzzy, few error links will affect the structure
greatly and the random factors of algorithm  will also induce a big
change in community structure. So, the only problem is how to
evaluate the effect of error links for community structure. We will
propose a method to evaluate the significance. The method admits
solid mathematical basis, so that the analysis of significance is
easy and reliable. Hence, the significance of community structure
can be evaluated effectively.

\subsection{Robustness of Community Structure}

We begin by defining the adjacency matrix $\textbf{A}$ of a network,
which consists of elements: $A_{i,j}=1$ when there is an edge
joining vertices $i$ and $j$; 0 otherwise. The corresponding
Laplacian matrix $\textbf{L}$ is defined as: $L_{i,j}=-A_{i,j}$ if
$i\neq j$, and $L_{i,i}=k_i$, where $k_i$ is the degree of node $i$.
$\lambda_{i}$ is the eigenvalue and $\textbf{v}_i$ is the
corresponding eigenvector of $\textbf{L}$. Moreover, we let
$0=\lambda_{1}\leq\lambda_{2}\leq\lambda_{3}\leq\cdots\leq\lambda_{n}$,
$\textbf{v}_i^{T}\textbf{v}_j=0$ if $i\neq j$, and
$\textbf{v}_i^{T}\textbf{v}_i=1$ for all $i$. In the well studied
bi-community problem \cite{newman_spectra} (partition the network
into two communities with pre-knowledge the size of each community),
the community structure vector $\textbf{s}$ with elements $s_i$ is
defined as: $s_i=1$ if node $i$ belongs to community $1$ and
$s_i=-1$ if node $i$ belongs to community $2$. $\textbf{s}$ can be
written as a linear combination of the normalized eigenvectors
$\textbf{v}_i$. Thus,
$\textbf{s}=\sum_{i=1}^{n}a_{i}\textbf{v}_{i}$, where
$a_i=\textbf{v}_i^T\textbf{s}$. Since $\textbf{s}^T\textbf{s}=n$,
$\sum{a_{i}^{2}}=n$, the bi-community problem can be written as an
optimization problem:
\begin{equation} Min Z=\textbf{s}^{T}\textbf{L}\textbf{s}=\sum{a_{i}^{2}
\lambda_{i}}.\label{bi-partation}\end{equation} where $\frac{1}{4}Z$
is the number of links between the two partitioned communities.

To minimize $Z$ is always a tough problem and can be equated with
the task of choosing the nonnegative quantities $a_i^2$ so as to
place as much as possible of the weight in the sum in the terms
corresponding to the lowest eigenvalues and as little as possible in
the terms corresponding to the highest eigenvalues
\cite{newman_spectra}. So the above optimization problem can be
simplified as:
\begin{equation} Min Z\approx Max \hat{Z}= a_{2}^{2}\lambda_2\end{equation}

Now we will extend the above bi-community network problem to
multi-community network one. Suppose that a network has $n$ nodes
and $c$ communities, and we have
$0=\lambda_{1}\leq\lambda_{2}\approx\lambda_{3}\approx\cdots\approx\lambda_{c}\leq\lambda_{c+1}\leq
\cdots \leq\lambda_{n}$ \cite{Systematic}. $\textbf{S}_1$ denotes
the community vector of community one. If node $i$ belongs to
community one, $S_{1,i}=1$ and $-1$ otherwise. Then
$\frac{1}{4}\textbf{S}_{1}^{T}\textbf{L}\textbf{S}_{1}$ is the
number of edges between community 1 and the rest of the network.
Consequently, we can define quantitatively the optimal partition
as:\begin{equation} Min
Z=\sum_{i=1}^{c}\textbf{S}_{i}^{T}\textbf{L}\textbf{S}_{i}.\end{equation}
Let
$\textbf{S}=(\textbf{S}_{1}^{T},\textbf{S}_{2}^{T},\cdots,\textbf{S}_{c}^{T})^{T}$
and
$\hat{\textbf{L}}=diag(\textbf{L},\textbf{L},\cdots,\textbf{L})$,
thus, we have
\begin{equation}Min Z=\textbf{S}^{T}\hat{\textbf{L}}\textbf{S}.\end{equation}

We can obtain all orthogonal and normalized eigenvectors
$\textbf{u}_q$ and the corresponding eigenvalues $\tau_q$ of
$\hat{\textbf{L}}$, where $q=1,2,\cdots,n\times c$. Obviously, each
eigenvalue of $\textbf{L}$ is $\hat{\textbf{L}}$'s eigenvalue and
repeat $c$ times. Without loss of generality, we let
$\tau_{ci-c+j}=\lambda_i, j=1,2,\cdots,c$. Let $\textbf{SU}$ be the
eigenvectors set of the eigenvalues of
$\lambda_2,\lambda_3,\cdots,\lambda_c$ of matrix $\hat{\textbf{L}}$.
$\textbf{SU}$ can be written as
$\textbf{SU}=\{(\textbf{v}_2^T,0\cdots,0),\cdots,(\textbf{v}_c^T,0,\cdots,0),\cdots,
(0,0,\cdots,\textbf{v}_c^T)\}$, where each $0$ denote an
$n$-dimensional zero vector and $\textbf{SU}$ has $c\times (c-1)$
elements. We can expand $\textbf{SU}$ as a space $\textbf{SSU}$ in
which each point is the liner combination of the elements in set
$\textbf{SU}$. The multi-partition problem can be written
as:\begin{equation}Min Z=\sum_{q=1}^{n\times
c}b_{q}^{2}\tau_q\approx Max \hat{Z}= \sum_{\textbf{u}_q\in
\textbf{SSU}}b_{q}^{2}\tau_q\approx\bar{\lambda}\sum_{\textbf{u}_q\in
\textbf{SSU}}b_{q}^{2}\end{equation} where
$b_{q}=\textbf{S}^{T}\textbf{u}_q$ and $\bar{\lambda}$ is the
average value of $\tau_{c+1}$ to $\tau_{c\times c}$ (also is the
average value of $\lambda_2$ to $\lambda_c$). $\sum_{\textbf{u}_q\in
\textbf{SSU}}b_{q}^{2}$ denotes the length of vector $\textbf{S}$
projection in space $\textbf{SSU}$. Obviously, the longer the
projection is, the nearer $\textbf{S}$ approaches the optimal. It is
difficult to obtain the optimal $\textbf{S}$. In this paper, we
focus on how to evaluate the significance of community structure.
Could we avoid the tough problem and measure the community structure
significance? For a network with a clear community structure, even
if there are a few error links the community structure should be
change a little. In contrast, when its community structure is fuzzy,
a few error links or a slight perturbation will lead to a big change
in the community structure. This property should be reflected in
space $\textbf{SSU}$. That is, for the same change of links, if the
community structure is significant, the space $\textbf{SSU}$ will
change a little; otherwise it will change considerably. The space
$\textbf{SSU}$ is expanded by the simple combination of
$\textbf{v}_2,\textbf{v}_3,\cdots,\textbf{v}_c$; therefore, the
robustness of space $\textbf{SSU}$ equals the robustness of the
eigenvalues $\lambda_2,\lambda_3,\cdots,\lambda_c$ and eigenvectors
$\textbf{v}_2,\textbf{v}_3,\cdots,\textbf{v}_c$.

Suppose that, $\delta A$ is the perturbation links for the original
network. Then, we can write $\delta L,\,\,\delta\lambda_i$ and
$\delta v_i$ as the corresponding perturbation of the Laplacian
matrix $L$ and its eigenvalues and eigenvectors. According to the
eigenvalue and eigenvector stability theory \cite{Russia}, we have
the following equations:
\begin{equation}(\delta \textbf{L}+\textbf{L})(\delta
\textbf{v}_i+\textbf{v}_i)=(\delta \lambda_i+\lambda_i)(\delta
\textbf{v}_i+\textbf{v}_i)\end{equation} by deleting the
second-order small quantities, we have
\begin{equation}\delta \textbf{L}\textbf{v}_i+\textbf{L}\delta \textbf{v}_i=\lambda_i\delta \textbf{v}_i+\delta \lambda_i \textbf{v}_i\end{equation}
after some deductions we obtain:
\begin{equation}\delta \lambda_i=\frac{\textbf{v}_{i}^{T}\delta \textbf{L}\textbf{v}_i}{\textbf{v}_{i}^{T}\textbf{v}_i},\,\,\,\, \delta \textbf{v}_i=\sum_{j=1}^{n}h_{ij}\textbf{v}_{j}\end{equation} where,
$h_{ij}=\frac{\textbf{v}_{j}^{T}\delta \textbf{L}
\textbf{v}_i}{\textbf{v}_{j}^{T}\textbf{v}_{j}(\lambda_i-\lambda_j)},
(i\neq j).$ Therefore, we have
\begin{equation}|\delta\lambda_i|\leq\|\delta \textbf{L}\|\end{equation}
which implies that for any network, no matter the community
structure is significant or not, the eigenvalues are only related to
the perturbation strength. In this way, the eigenvalues are always
stable \cite{Russia}. (So, it is not necessary to consider the
stability of eigenvalues.)

Without loss of generality, we can let $a_{i1}=a_{ii}=0$ for $i\neq
1$. Then the comparative error of $v_i$ can be denoted as
\begin{equation}\frac{|\delta
\textbf{v}_i|}{|\textbf{v}_i|}\leq\parallel\delta
\textbf{L}\parallel\sum_{j\neq
i,j=2}^{n}\frac{1}{|\lambda_i-\lambda_j|}\label{single_lambda}\end{equation}
In Eq.\ref{single_lambda}, $\parallel\delta L\parallel$ is the
perturbation strength and $\sum_{j\neq
i,j=2}^{n}\frac{1}{|\lambda_i-\lambda_j|}$ is the amplification
coefficient which is used to measure the stability of
$\textbf{v}_i$. Integrating the stability of $\lambda_2$ to
$\lambda_c$, we define $R$ as the stability index of space
$\textbf{SSU}$.
\begin{equation}R=\sum_{j=c+1}^{n}\frac{1}{|\bar{\lambda}-\lambda_j|}\label{Rob}\end{equation}
Of cause, $R$ is an important index of the network which can be used
to measure the significance of community structure.

\subsection{Index of the Significance}

Although $R$ makes sense mathematically, it is not convenient to
measure and further compare the significance of different networks.
In this section, we will define an efficient index to measure the
community structure significance. Like the definition of
temperature, if we know the most significant and fuzzy stability
values $R$, the robustness can be scaled into interval $[0,1]$ which
will be very intuitive to use.

What kind of network possess the most significant community
structure? Suppose that the network size is $n$, the average degree
is $k$ and the community number is $c$,  where $c<<n$. To find the
most significant community structure is to solve the following
optimization problem:
\begin{equation} \label{eq:1}
\left\{ \begin{aligned}
         Min R=\sum_{i=c+1}^{n}\frac{1}{\lambda_{c+1}-\bar{\lambda}} \\
         s.t. \ \ \ \ \ \ \ \  \ \ \sum_{i=1}^{n}\lambda_{i}=nk.
                          \end{aligned} \right.
                          \end{equation}

%\begin{equation}Min R=\sum_{i=c+1}^{n}\frac{1}{\lambda_{c+1}-\bar{\lambda}},\end{equation}
%and the constraint is
%\begin{equation}\sum_{i=1}^{n}\lambda_{i}=nk.\end{equation}

\begin{figure}
\center
\includegraphics[width=4.4cm,height=3.8cm]{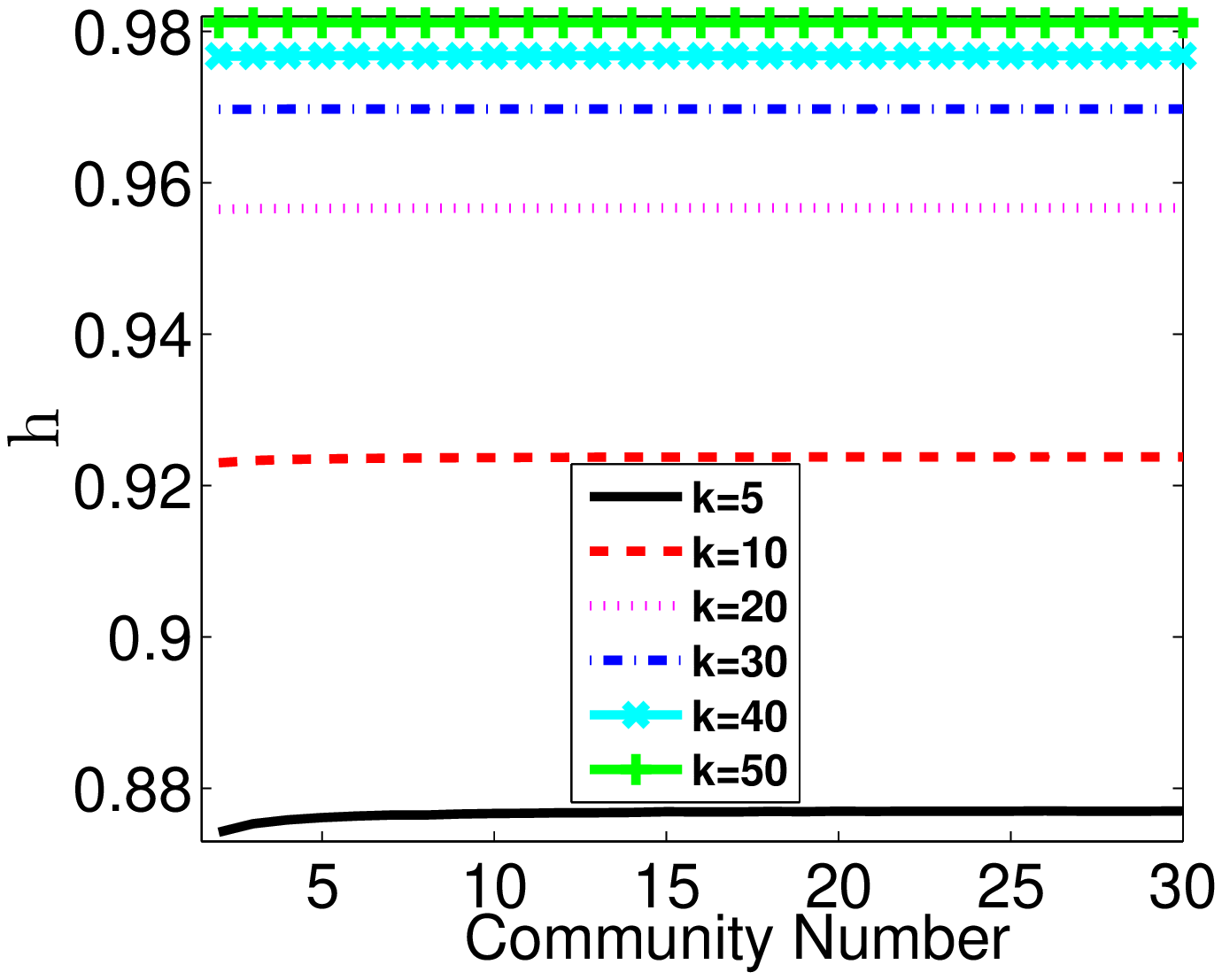}\includegraphics[width=4.4cm,height=3.8cm]{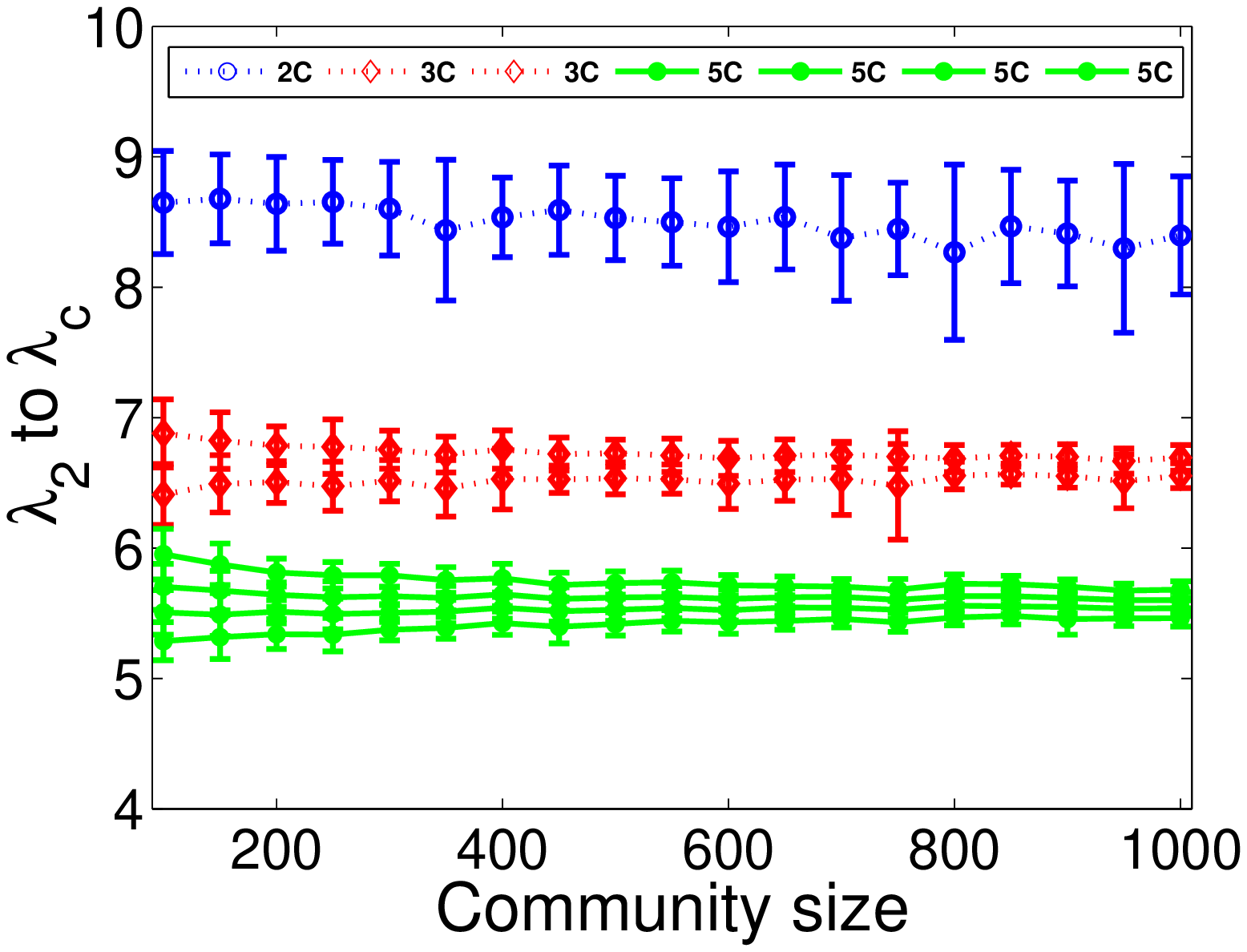}
\caption{\textbf{a.} The dependence of maximum $h$ and community
number on artificial connected networks. Given a degree $k$, each
community is a completely connected subgraph with $k+1$ nodes. Among
the communities there are $c-1$ connections making the whole network
connected, where $c$ is the number of communities. From the plot we
can see that the maximum $h$ is very close to 1. \textbf{b.} The
dependence of $\lambda_2$ to $\lambda_c$ and community size. In this
plot, $2C,3C,5C$ denote that there are 2,3 and 5 equal communities
in the network respectively. Each node has 20 expected links with
its fellows with the same community and 5 expected links with other
communities. The error bar denotes the standard deviation. We can
see that the standard deviation of $\lambda_2$ to $\lambda_c$ is
small and it does not depend on the size of community
considerably.}\label{MinRnofluction}
\end{figure}

For the above optimization problem, we directly set
$\bar{\lambda}=0$. By the Lagrange multiplier method, we obtain that
when $\bar{\lambda}=0, \lambda_{c+1},
\lambda_{c+2},\cdots,\lambda_n=\frac{nk}{n-c}$, $R$ will achieve
it's global minimum value $R=\frac{(n-c)^2}{nk}\approx\frac{n}{k}$.
$\bar{\lambda}=0$ implies that there are no any connections among
communities and the network is not connected which is not suitable
for our basic assumption. But this kind of unconnected network can
be modified slightly to meet our requirement. We can generate a
network with $c=\frac{n}{k+1}$ communities, and each community,
which is a completely connected subgraph, contains $k+1$ nodes.
Among the $c$ communities there are only $c-1$, connections which
guarantee that the whole network is connected. For this kind of
network, $\bar{\lambda}\approx0, \lambda_{c+1},
\lambda_{c+2},\cdots,\lambda_n\approx k+1$, and the corresponding
$R$ will achieve the global minimum value $R\approx\frac{n}{k}$, as
shown in Fig. \ref{MinRnofluction} \textbf{a}.

\begin{figure}
\center
\includegraphics[width=4.4cm,height=3.8cm]{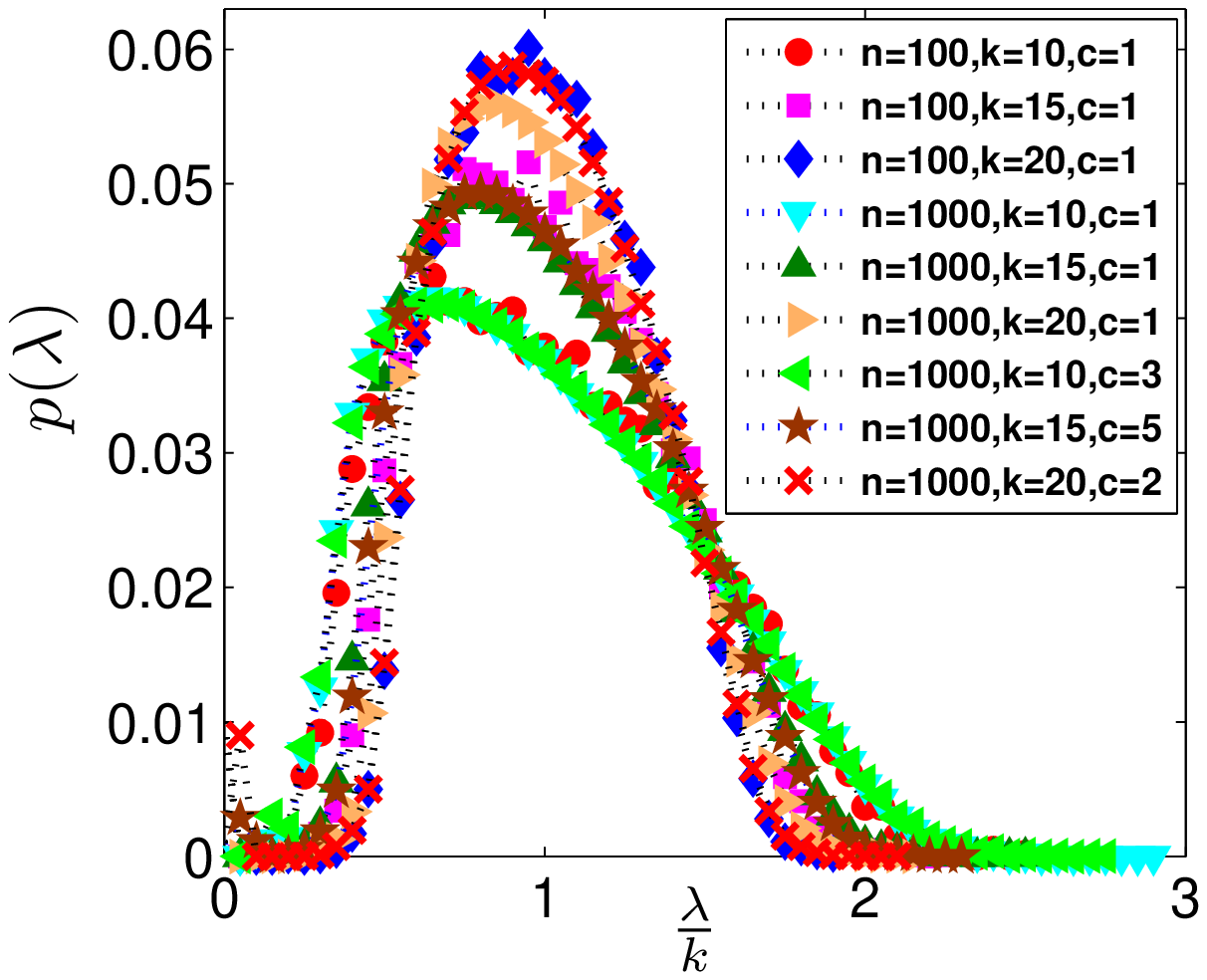}\includegraphics[width=4.4cm,height=3.8cm]{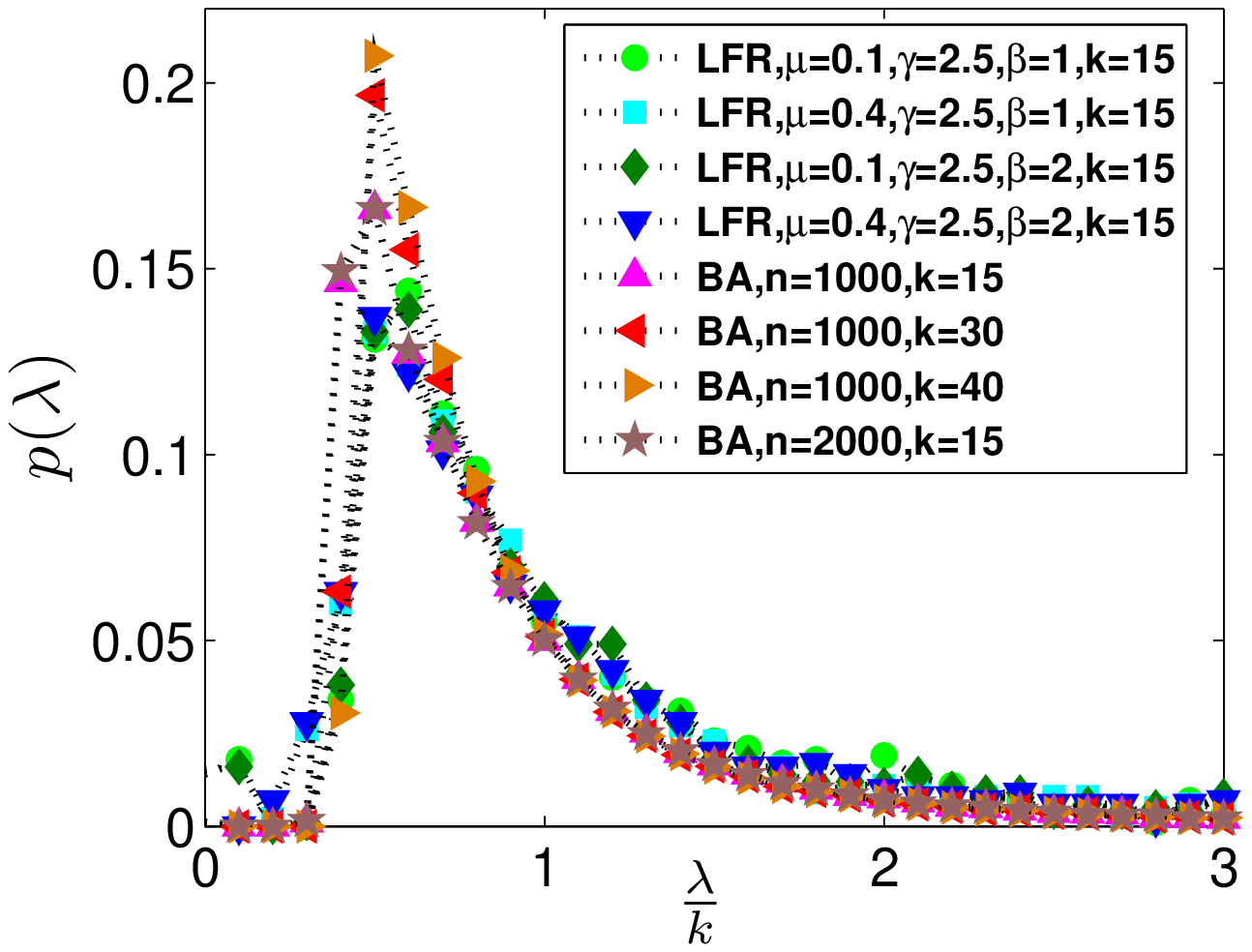}
\caption{The distribution of laplace matrix's eigenvalues. Community
structure has no considerable impact on its eigenvalue's
distribution for both homogeneous and heterogeneous networks. The
eigenvalues' distribution can be scaled by average degree for both
homogeneous and heterogeneous networks. The relative width of this
peak decreases with increasing average degree, while other
qualitative features are the same.\textbf{a.} In homogeneous
networks, each community is a ER network. The legend $c=1$ means
that the network is a ER network without communities. When $c=2$,
the two communities size is $100$ and $ 900$ respectively. When
$c=3$, there are 3 communities and the communities size is $100,
200, 700$. $c=5$, all the community sizes are 200. We also test many
other community size distribution and the results are same (not yet
been shown here). \textbf{b.} We employ the LFR-benchmark and the BA
model to generate the heterogeneous network. In the LFR-benchmark,
we set the maximum degree as $50$, and the maximum and minimum
community sizes is $50$ and $20$ respectively.}\label{eigdist}
\end{figure}

The spectra properties of complex network matrix have been well
studied \cite{Num_spec,Laplacian_spectra}. They throw a light on the
universal properties of the eigenvalues' distribution of random
spares matrices. We investigate the distribution of eigenvalues for
different community structures in both homogeneous (passion) and
heterogeneous (scale free) degree distribution networks. The results
show that the distribution of eigenvalues is  mainly  determined by
the average degree and degree distribution, and does not relate to
the community structure considerably (as shown in Fig.
\ref{eigdist}). Moreover, for networks with different size and
community structure, we investigate the most relavent eigenvalues
$\lambda_2, \lambda_3,\cdots,\lambda_c$, and we also find that they,
staying, depend only on the community structure and does not related
to the size of both community and network (as shown in Fig.
\ref{MinRnofluction} \textbf{b}).

\begin{figure}
\center
\includegraphics[width=4.4cm,height=3.8cm]{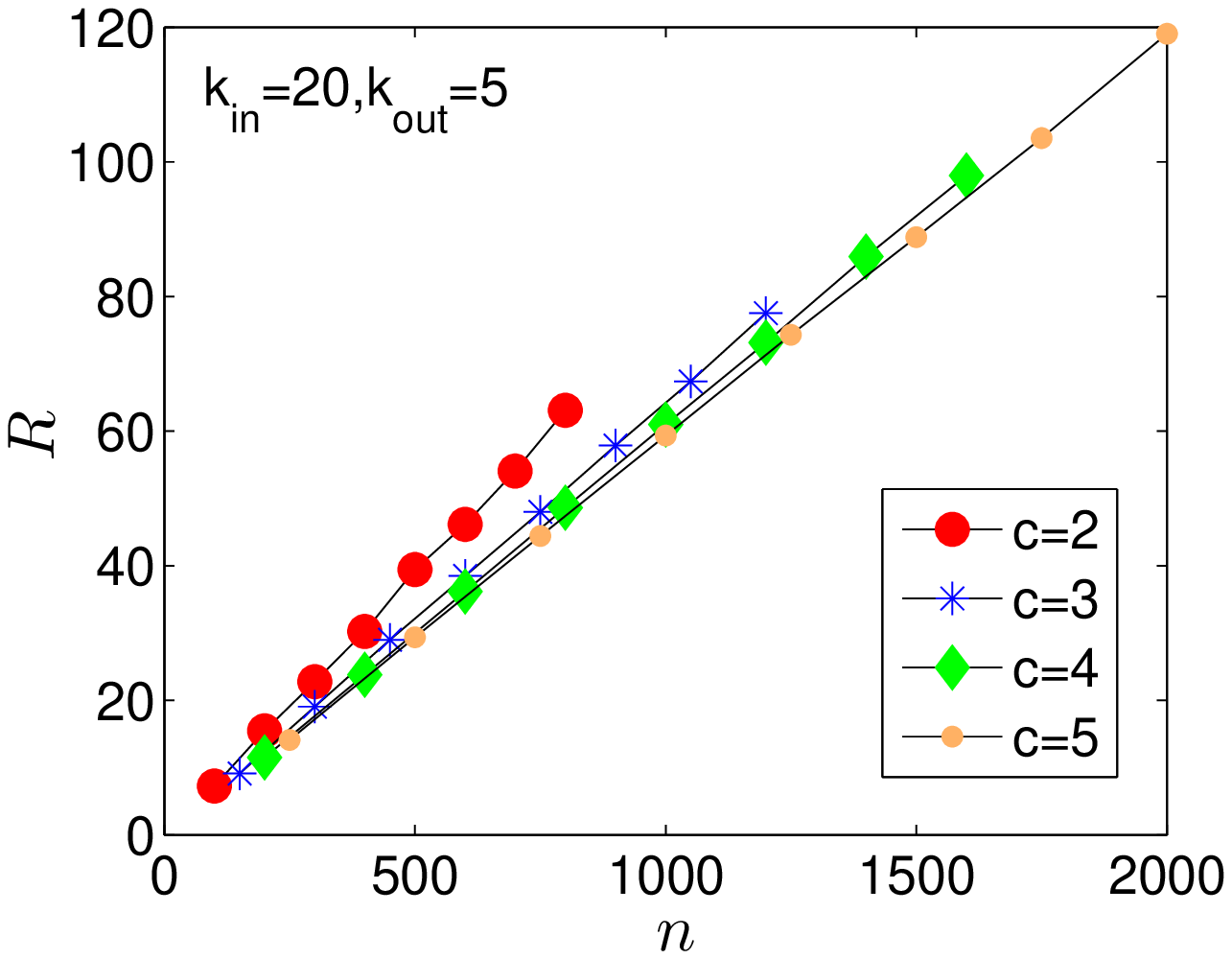}\includegraphics[width=4.4cm,height=3.8cm]{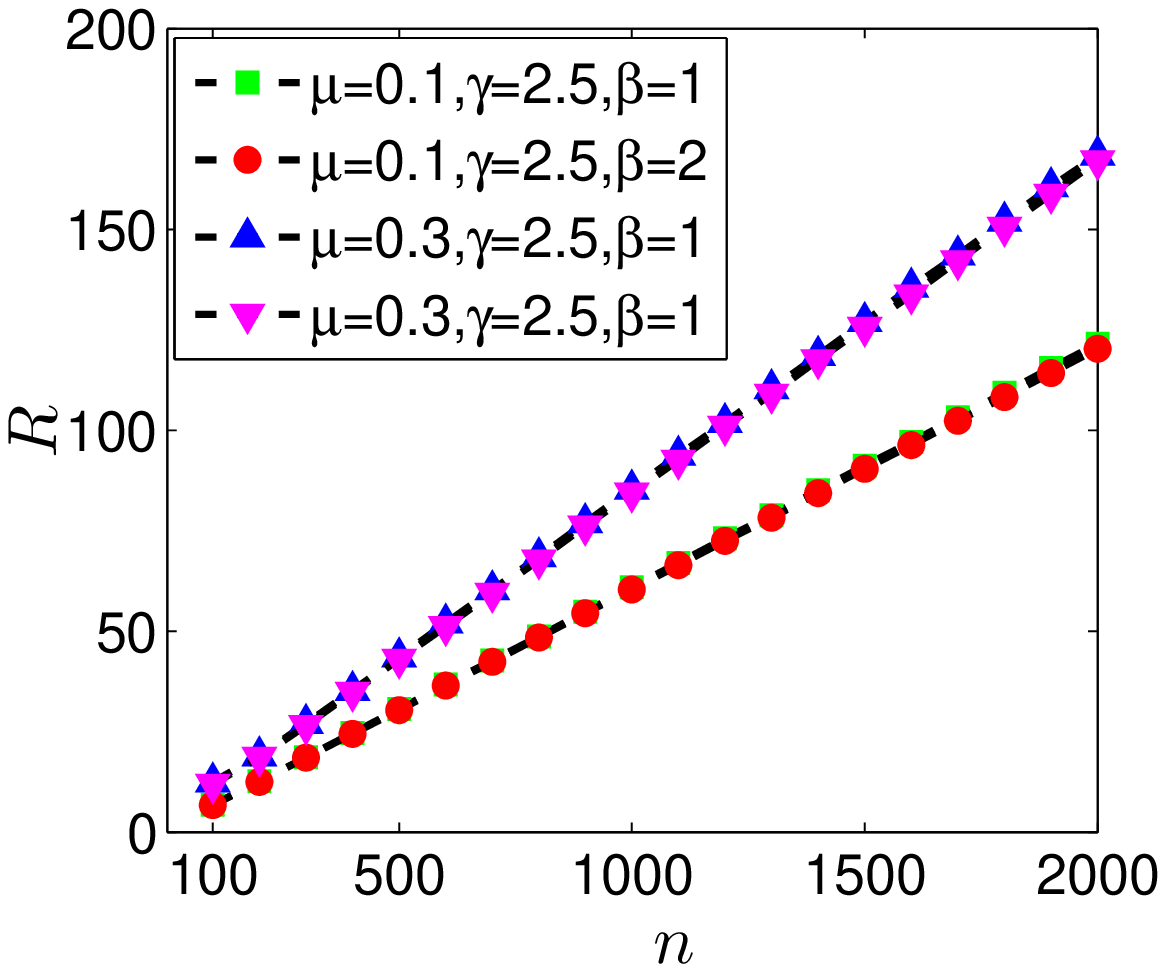}
\includegraphics[width=4.4cm,height=3.8cm]{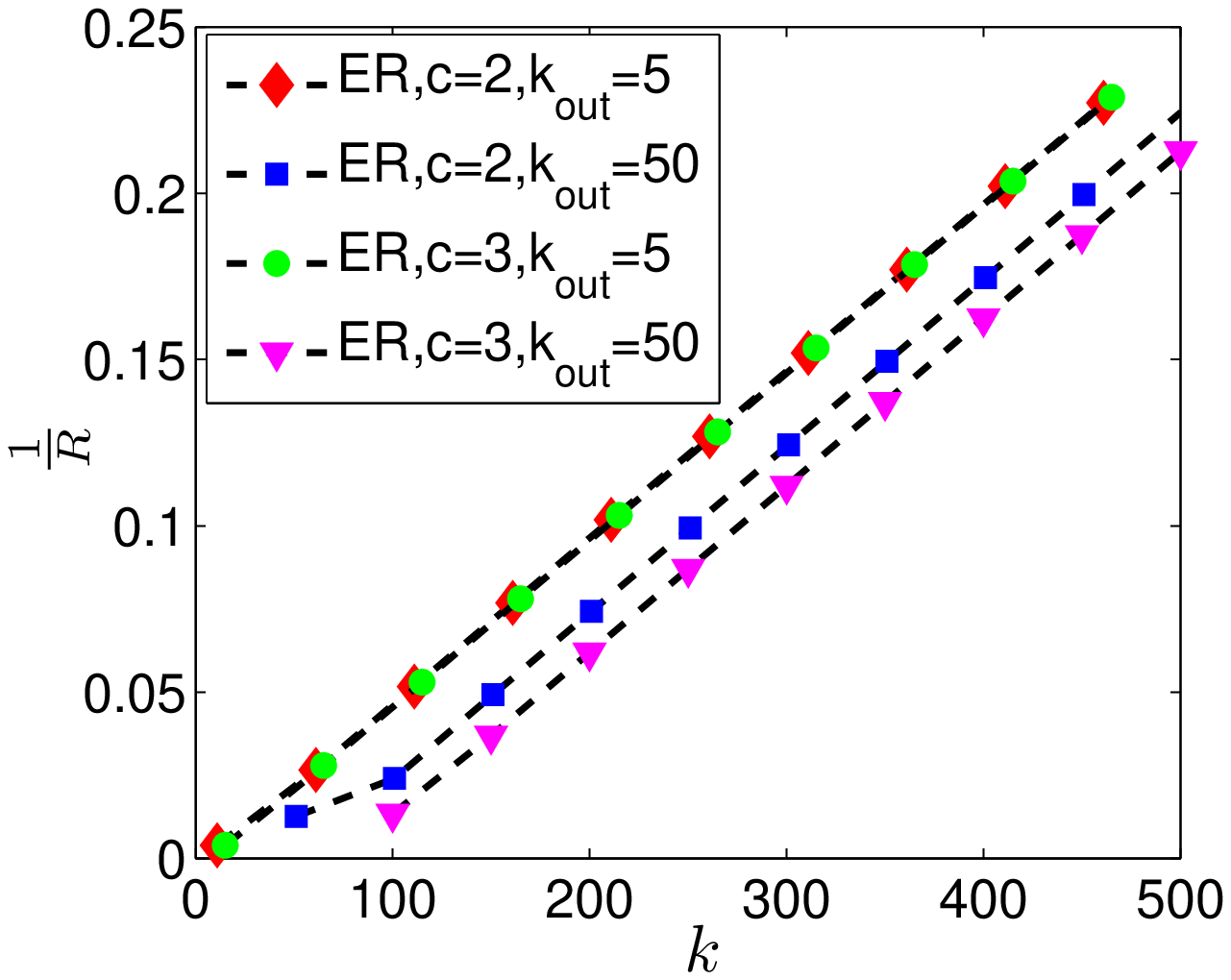}\includegraphics[width=4.4cm,height=3.8cm]{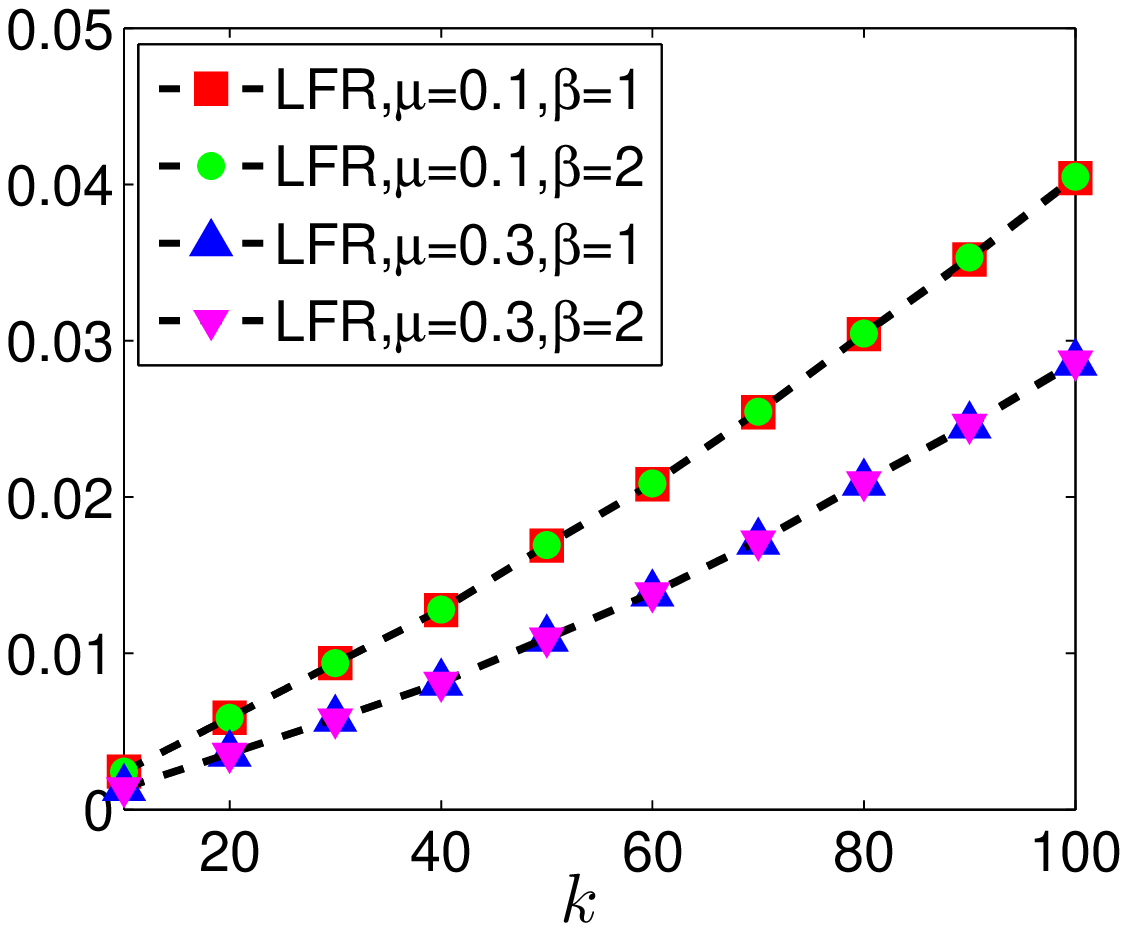}
\vskip -23.5em \hskip -27.0em  \raisebox{1em}{\large \bf{a}} \vskip
-1.9em \hskip 1.8em {\large \bf{b}} \vskip 10.5em \hskip
-26.5em{\large \bf{c}} \vskip -1.0em \hskip 1.8em {\large
\bf{d}}\vskip 9.0em \hskip -20.0em \caption{The relationship of
$R\propto\frac{n}{k}$. From the four plots we can see that
$R\propto\frac{n}{k}$ is almost validate. \textbf{a.} The dependence
of $R$ on the network size $n$. Each community size is the same and
each node has 20 links to others in the same community and 5 links
to the outside nodes. $c$ is the community number. \textbf{b.} The
dependence of $R$ and network size $n$ in the LFR benchmark. The
average degree $k=20$ and $P(k)\propto k^{2.5}$, the maximum degree
is $50$, maximum and minimum community size is $20, 50$
respectively. \textbf{c.} The dependence of $\frac{1}{R}$ and
average degree $k$. For different network size $n$, each community
size are same. \textbf{d.} The dependence of $\frac{1}{R}$ and
average degree $k$ in the LFR benchmark.The maximum degree is $200$
and maximum and minimum community size is $200, 100$ respectively.
Average degree is 20 and $P(k)\propto k^{2.5}$.}\label{linear}
\end{figure}

According to \cite{Num_spec,Laplacian_spectra}, Eq. \ref{Rob} and
Fig.\ref{eigdist}, we have $R\propto n$ strictly as shown in Fig.
\ref{linear}. From Fig. \ref{linear} we can see that, for both
homogeneous and heterogeneous degree distribution, the great mass of
eigenvalues $\lambda$ are near the average degree, although the
distribution of eigenvalues can not be scaled by the average degree.
We have conducted many numerical experiments in both homogeneous and
heterogeneous networks and find that $\frac{1}{R}\propto k$ holds
well. % (although not strictly, shows to hold good.)
Therefor, given a network with robustness $R=h\frac{n}{k}$, when the
community structure is more significant, $h$ will be small. From the
above analysis of the clearest community structure in a large enough
network, we have a lower bound, which almost approaches $1$. It is
very hard to get the $h$ of a fuzziest community structure for that
the continuous property of matrix spectra is very complicated. So,
to simplify the index, we define $H=\frac{1}{h}=\frac{n}{Rk}$ as the
significance of community structure and $H$ is almost in $[0,1]$
when network size is large enough.

\section{Result}
\subsection{Artificial Networks}
\begin{figure}
\center
\includegraphics[width=4.4cm,height=4.2cm]{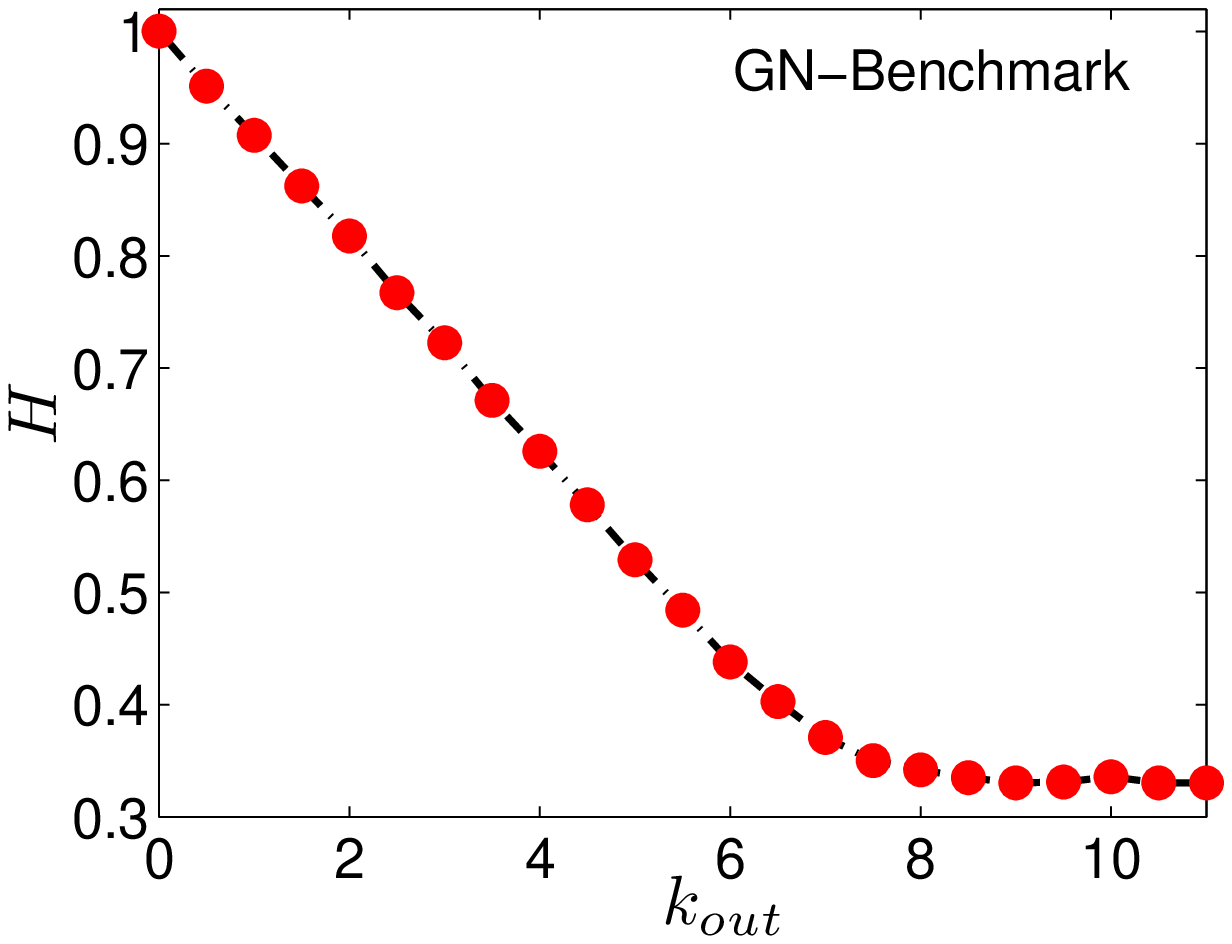}\includegraphics[width=4.4cm,height=3.8cm]{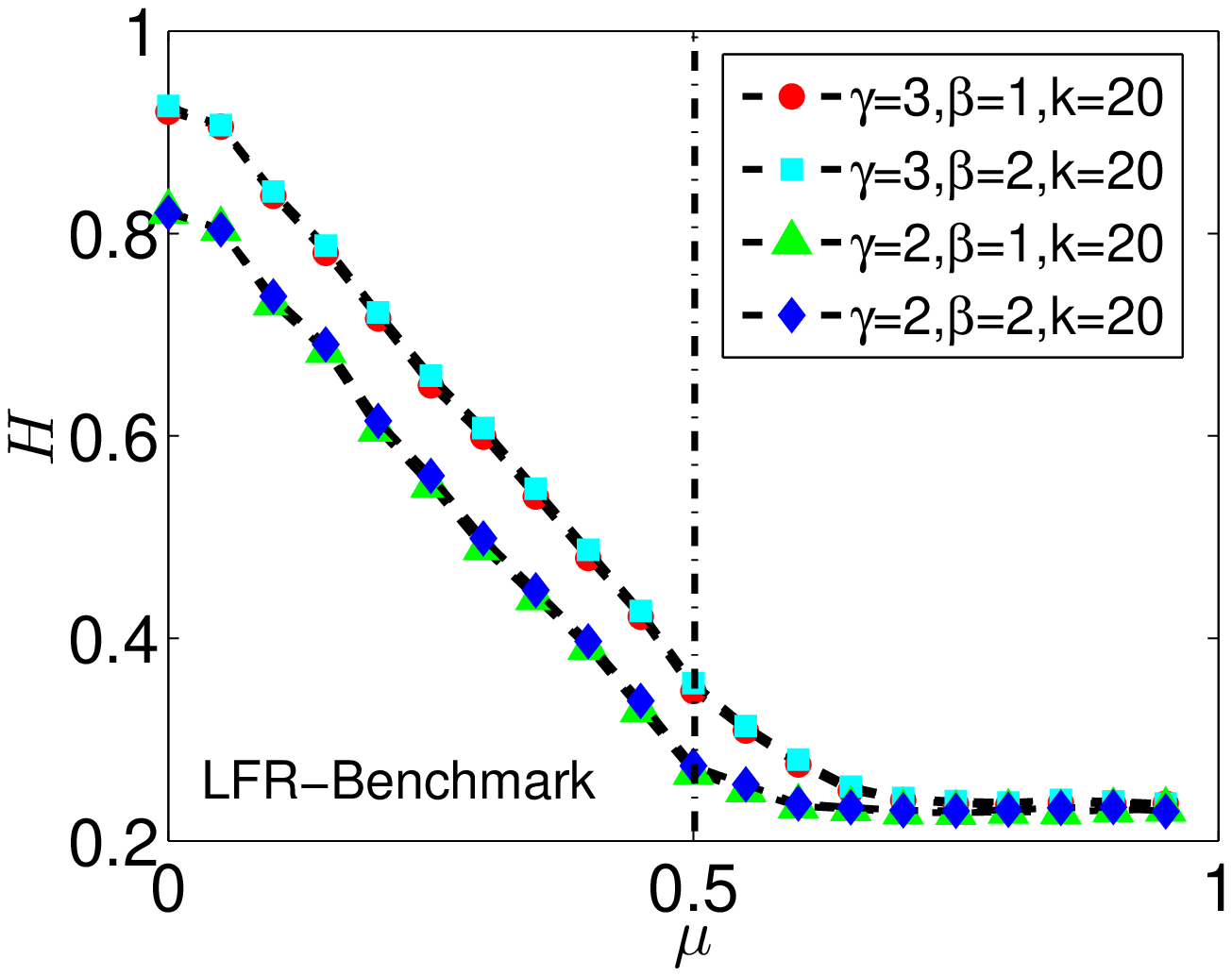}
\caption{The performance of $H$ index in both GN-benchmark and
LFR-benchmark. In GN-benchmark, we can see that $H$ decrease with
increasing of $k_{out}$. When the community structure is very clear
$H$ close to 1 very much, and the network close to no community
structure network  $H$ close to 0.3 which implies that for a given
network when $H$ is less than 0.3 it is not safe to say there exit
significant community structure. In LFR-benchmark, the average
degree $k=20$, maximum degree is $50$ and $P(k)\propto k^{\gamma}.$
Maximum and minimum community sizes are $50$ and $20$ respectively,
more over $Q(m)\propto m^{\beta}$ where, $m$ denotes the community
size. We can see that with the increase of mix parameter $\mu$, the
$H$ index decrease. When $\mu\geq0.5$ (no significant community) $H$
is near $0.3$ which is similar with GN-benchmark.}\label{benchmark}
\end{figure}

Let's test the validity of our index. Firstly, we use the classical
GN benchmark presented by Girvens and Newman \cite{Newman_Q}. Each
network has $n=128$ nodes that are divided into $4$ communities with
$32$ nodes each. Edges between two nodes are introduced with
different probabilities which depend on whether the two nodes belong
to the same community or not. Each node has $\langle k_{in}\rangle$
links on average with its fellows in the same community, and
$\langle k_{out}\rangle$ links with the other communities, and we
keep $\langle k_{in}\rangle+\langle k_{out}\rangle=16$. As is well
known, the communities become fuzzier and thus more difficult to be
identified when $k_{out}$ increases. Hence, the significance of the
community structure will also tend to be weaker and the $R$ index
will decrease. The numerical experiments' results are shown in Fig.
\ref{benchmark}. We can find that the index $H$ works well in the
GN-benchmark. When community structure is very clear, the $H$ is
very close to $1$; when the network is nearly a random one, the
corresponding $H$ is near to $0.3$. Thus, we argue that for a given
network when the corresponding $H$ is larger than 0.3, there exists
community structure. Moreover, the larger the $H$ index is, the more
significant community structure will be.

We also test the index on the more challenging LRF benchmark
presented by Lancichinetti, Fortunato, Radicchi \cite{LFR}. In the
LFR benchmark, each node is given a degree took from a power law
distribution with an exponent $\gamma$, and the sizes of the
communities are took from a power law distribution with an exponent
$\beta$. Moreover, each node shares a fraction $1-\mu$ of its links
with other nodes of its community and a fraction $\mu$ with other
nodes in the network. $\mu$ is the mixing parameter. The community
structure significance can be adjusted by the mixing parameter
$\mu$. The numerical results in the LFR-benchmark are shown in Fig.
\ref{benchmark}. We can see that $H$ decreases with the augment of
$\mu$ and $H$ is independent of the community size distribution.
Moreover, when the power law exponent of degree distribution becomes
larger, the community structure will be more significant. That more
homogenous the degree distribution is, the more significant the
community structure will be, when other conditions are same.

\subsection{Real-world Networks}
\begin{figure}
\center

\includegraphics[width=4.4cm,height=3.8cm]{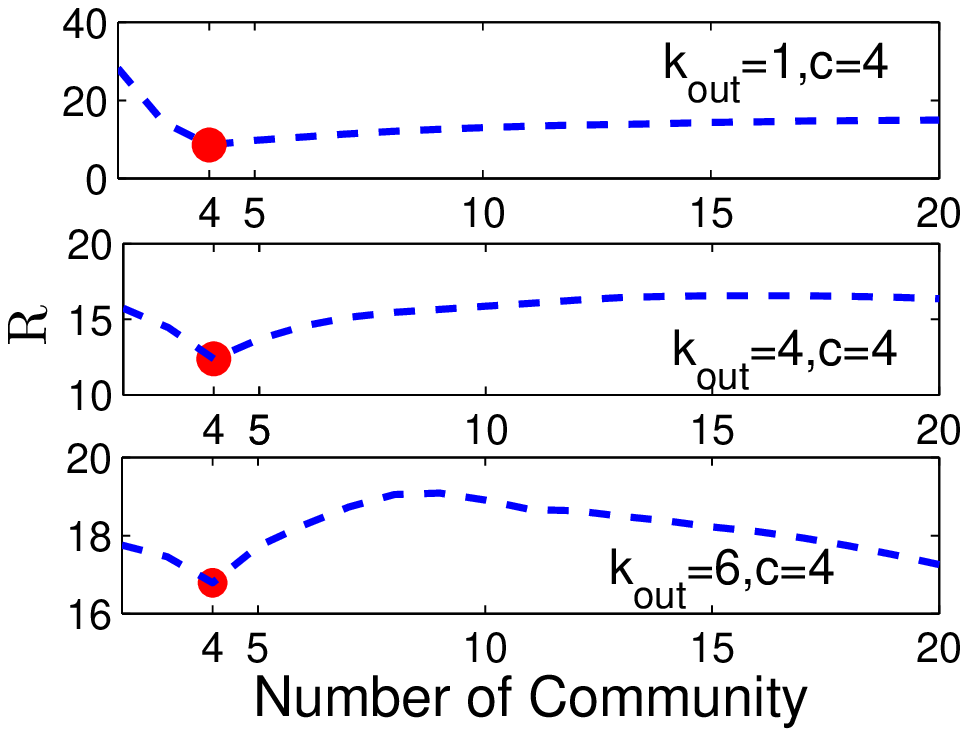}\includegraphics[width=4.4cm,height=3.8cm]{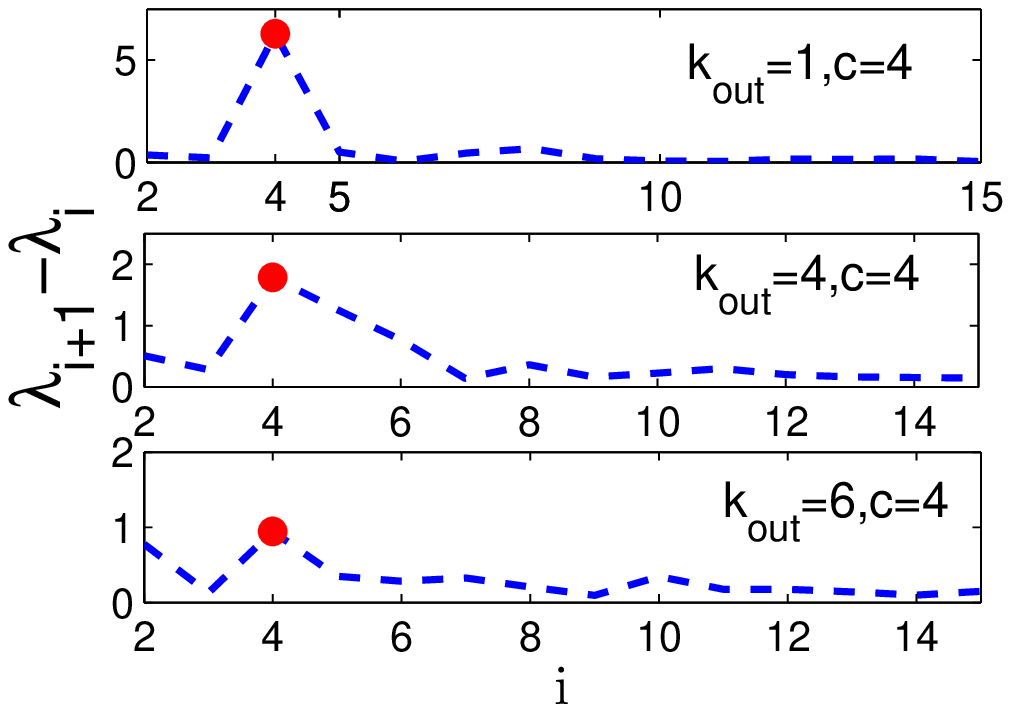}
\includegraphics[width=4.4cm,height=3.8cm]{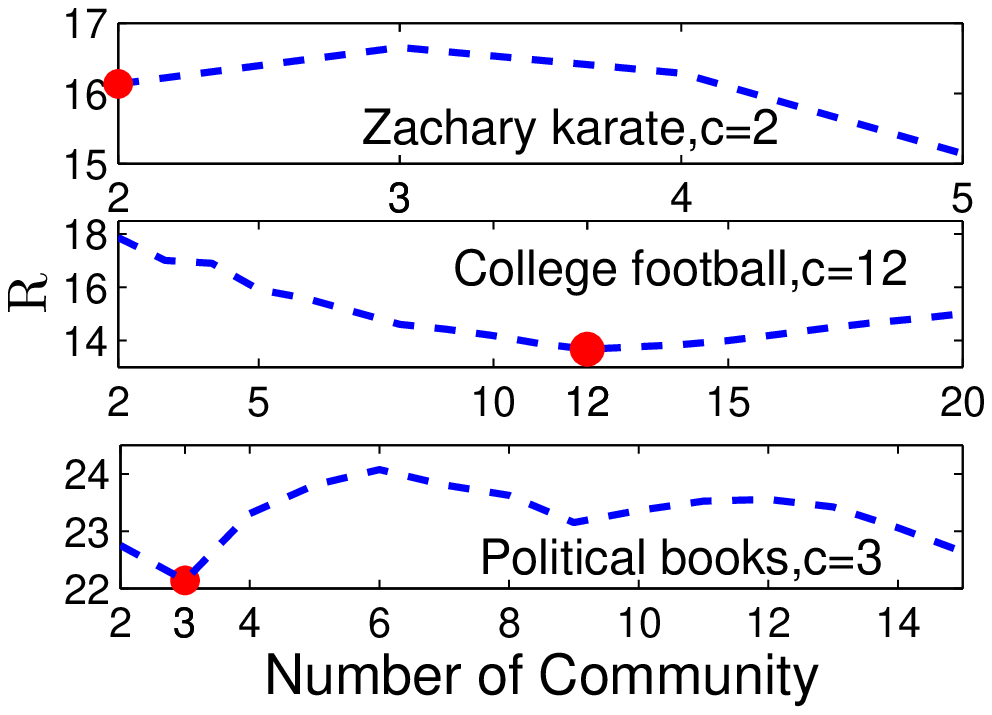}\includegraphics[width=4.4cm,height=3.8cm]{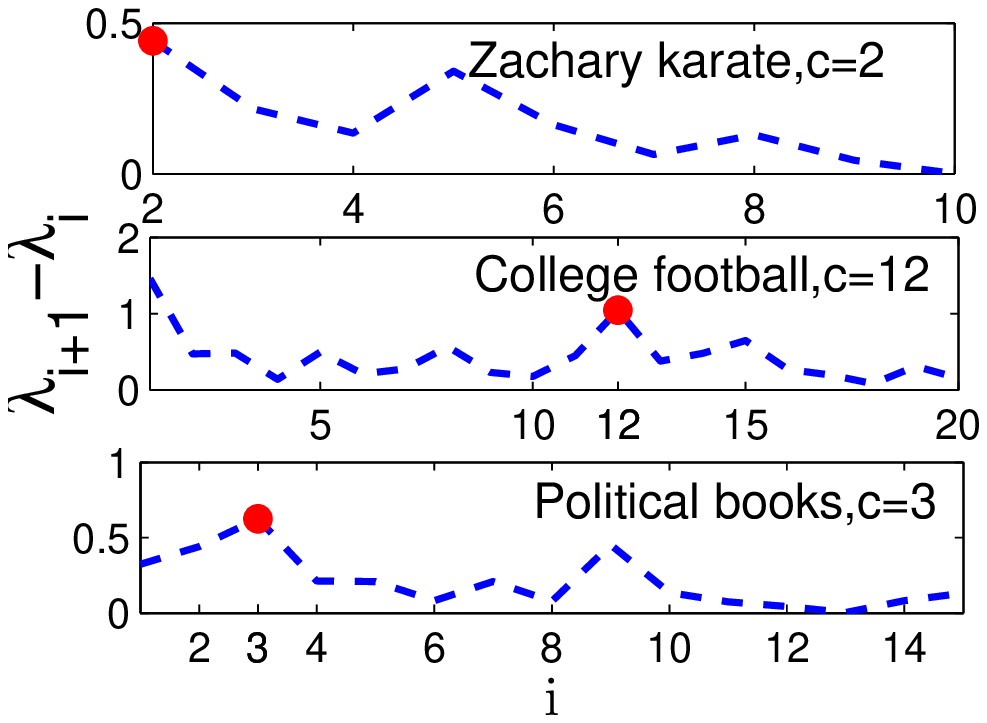}
\vskip -23.5em \hskip -27.0em  \raisebox{1em}{\large \bf{a}} \vskip
-1.9em \hskip 1.8em {\large \bf{b}} \vskip 10.5em \hskip
-26.5em{\large \bf{c}} \vskip -1.0em \hskip 1.8em {\large
\bf{d}}\vskip 9.0em \hskip -20.0em\caption{The optimal community
number. The rang optimal $c$ is about $\frac{n}{10}$ for all
networks. From the plots \textbf{a} and \textbf{b} we can find that
on GN-benchmark, when $c=4$ (pre-determined community number) $R$
achive lowest value and the corresponding $\lambda_{c+1}-\lambda_c$
also achieve the largest value. \textbf{c} and \textbf{d} The
empirical results of Zachary karate club network, College football
network and Political books network. It was found that Zachary
karate club has 2 communities and College football network has 12
communities. Form the plots we can see that $R$ achieve its lowest
value when the community numbers correspond the reality. The
corresponding values of $\lambda_{i+1}-\lambda_i$ also present
reasonable phenomenons. For the Political books, we still don't know
how many communities, the method shows that it has 3
communities.}\label{optimal-c}
\end{figure}

Till now, we still haven't discuss how to obtain the optimal
community number $c$. For many real-world networks, we don't know
the community number before calculating the index value or
partition. Many numerical experiments (as shown in Fig.
\ref{optimal-c}) support that the community structure will be most
clearest when the community number is the optimal $c$. So generally
speaking, the corresponding community number with the lowest $R$
will be the optimal $c$. Moreover, at the optimal $c$, the value of
$\lambda_{c+1}-\lambda_c$ will be very large comparatively. So we
also can resort to the differences between $\lambda_i$ and
$\lambda_{i+1}$ to detect the optimal $c$.

We apply the index to many real networks (see Tab.\ref{Table real
networks} and detail information in supplementary). The data are
taken from the following references and web sites
\cite{ZK_d,Dolphin,Ploblog,C.neural,Jazz,C.met,webset1,webset2,webset3}.
People usually classify the real networks into three categories:
social networks (such as scientist collaborations and friendships),
biological networks (such as proteins interaction networks and
metabolic networks) and technological networks (such as Internet and
the WWW). First, we analyze several social networks, including
Zachary karate club network \cite{ZK_d}, dolphin network
\cite{Dolphin}, collage football network \cite{GN}, Jazz network
\cite{Jazz}, scientists collaboration network \cite{webset1} and so
on. The results are very similar to our previous work \cite{Yhu}. We
find that the Jazz community structure is the most significant one,
the Santa Fe scientists collaboration network and the Political
blogs network are insignificant comparatively. Generally speaking,
the community structure is most notable in social networks.
Moreover, we analyze some biological networks such as proteins
interaction networks (E. coli \cite{webset2}, Yeast \cite{webset3}
and H. Sapiens \cite{webset2}), many metabolic networks
\cite{webset3} and C.elegans neural network. We find that in
proteins interaction networks, E.coli is 0.14, H. Sapiens 0.21, and
Yeast 0.40, which is high and different from the previous results.
In metabolic networks, the $H$ index of Aquifex aeolicus,
Helicobacter pylori and Yersinia pestis are all 0.36, which are
consistent to previous works. But for the C.elegans metabolic and
neural network , significance is 0.62, which is very high and
different from previous work due to it is not easy to obtain the
proper community number $c$ (see supplementary). The significance of
C.elegans neural is 0.57, which corresponds to previous work well.

%\begin{figure} \center
%\includegraphics[width=7cm]{500ER.eps}\includegraphics[width=7cm]{500BA.eps}
%\caption{}\label{}
%\end{figure}

\begin{table}
\caption{The $\hat{R}$ and $H$ indexes of some real networks.
$\hat{R}$ indexes is the robustness of community structure, which
can be obtain be perturbation (please see ref \cite{Yhu}). The table
shows the names of different real networks and the corresponding
index values.}\label{Table real networks}

\begin{tabular} [t]{|c|c|c|c|c|c|c|}
\hline network&size&$\hat{S}$&$\hat{R}$&H&type\\\hline
E.coli&1442, 5873&61.30&0.11&0.14&\\
Yeast&1458, 1971&112.95&0.12&0.40&protein\\
H.Sapiens&693, 982&38.48&0.18&0.21&\\\hline
C.elegans metabolic&453, 2032&19.25&0.17&0.62&\\
Aquifex aeolicus&1473, 3354&68.39&0.17&0.36&\\
Helicobacter pylori&1341, 3087&62.76&0.17&0.36&metabolic\\
Yersinia pestis&1922, 4389&108.84&0.15&0.36&\\
43 metabolic networks&1472, 3395&71.25&0.17&0.36&\\\hline C.elegans
neural&297, 2148&5.52&0.22&0.52&neural\\\hline
Santa Fe scientists&118, 200&2.45&0.27&0.72&\\
Zachary karate&34, 78&0.32&0.25&0.46&\\
Dolphin&62, 159&2.07&0.24&0.42&\\
College
football&115, 613&1.67&0.34&0.79&social\\
Jazz&198, 2742&0.64&0.40&0.47&\\
Email&1133,5452&27.35&0.19&0.42&\\
Political blogs&1222, 19090&0.57&0.27&0.22&\\
Political books&105, 441&1.63&0.31&0.32&\\\hline
\end{tabular}
\end{table}

\section{Conclusion and discussion}
In this paper, an index to evaluate the significance of community
structure without knowing the community structure is proposed. We
transform the problem of community structure significance into the
problem of the stability of eigenvalues and eigenvectors of the
Laplacian matrix. The index of community structure significance
admits sound mathematical basis, which makes the index is reliable.
According to the index, the optimal community number can also be
obtained before partition, which is nearly impossible for many
partition algorithms. Moreover, we apply the index to many real
world networks, such as social networks, neural network,
protein-interaction networks and metabolic networks. We find that in
social networks, the significance of community structure is usually
high, C.elegans metabolic and neural networks they are very hight,
and in protein interaction and some other metabolical, they are
comparative low.

\section*{Acknowledgement}
Yanqing Hu wishes to thank Prof. Shlomo Havlin for very useful
discussions, Dr. Erbo Zhao for his help in compiling LFR-benchmark
and Dan Bu for some help in English writing. This work is partially
supported by 985 Project and NSFC under the grant No. 70771011, and
No. 60534080.

\section{Supplementary}

\begin{figure}
\center
\includegraphics[width=8cm]{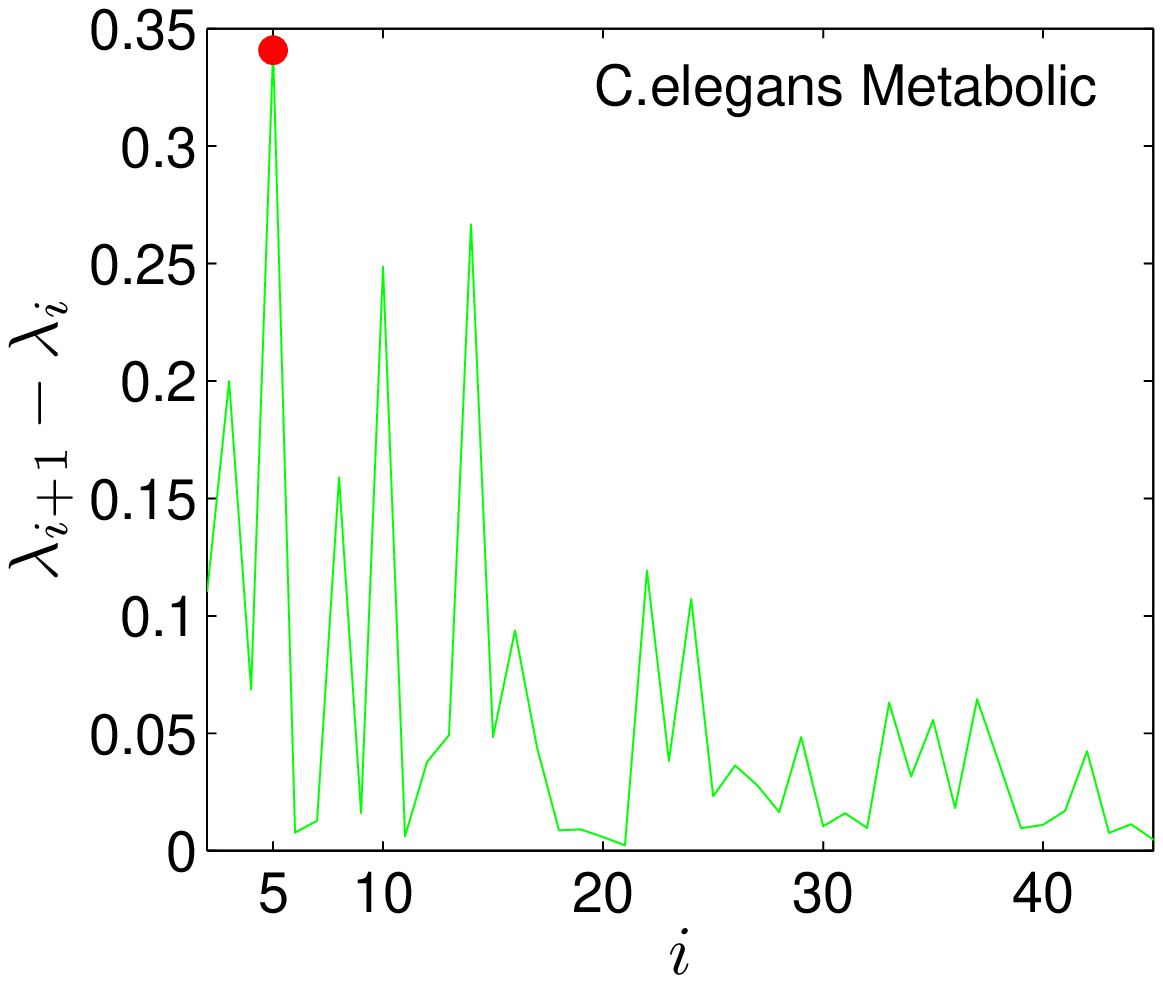}\includegraphics[width=8cm]{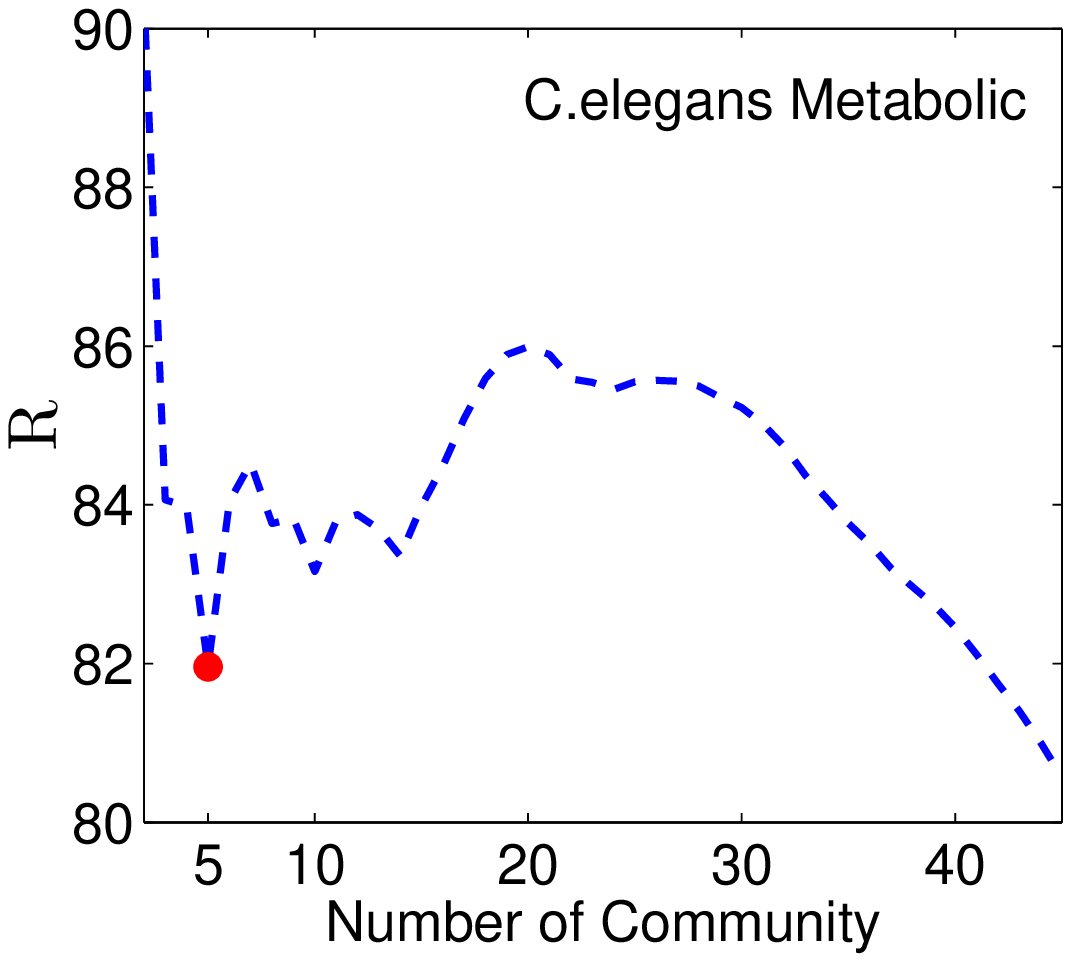}
\includegraphics[width=8cm]{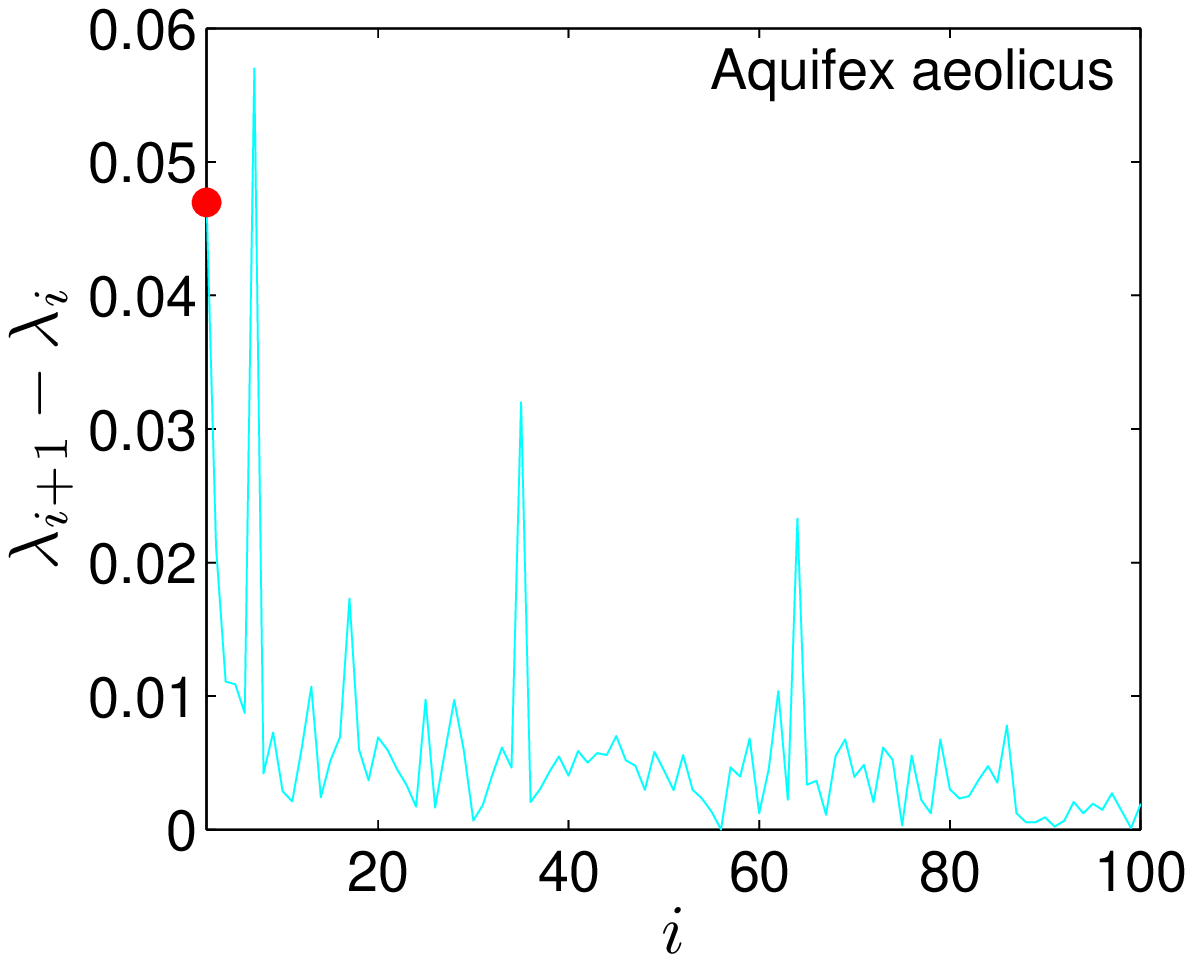}\includegraphics[width=8cm]{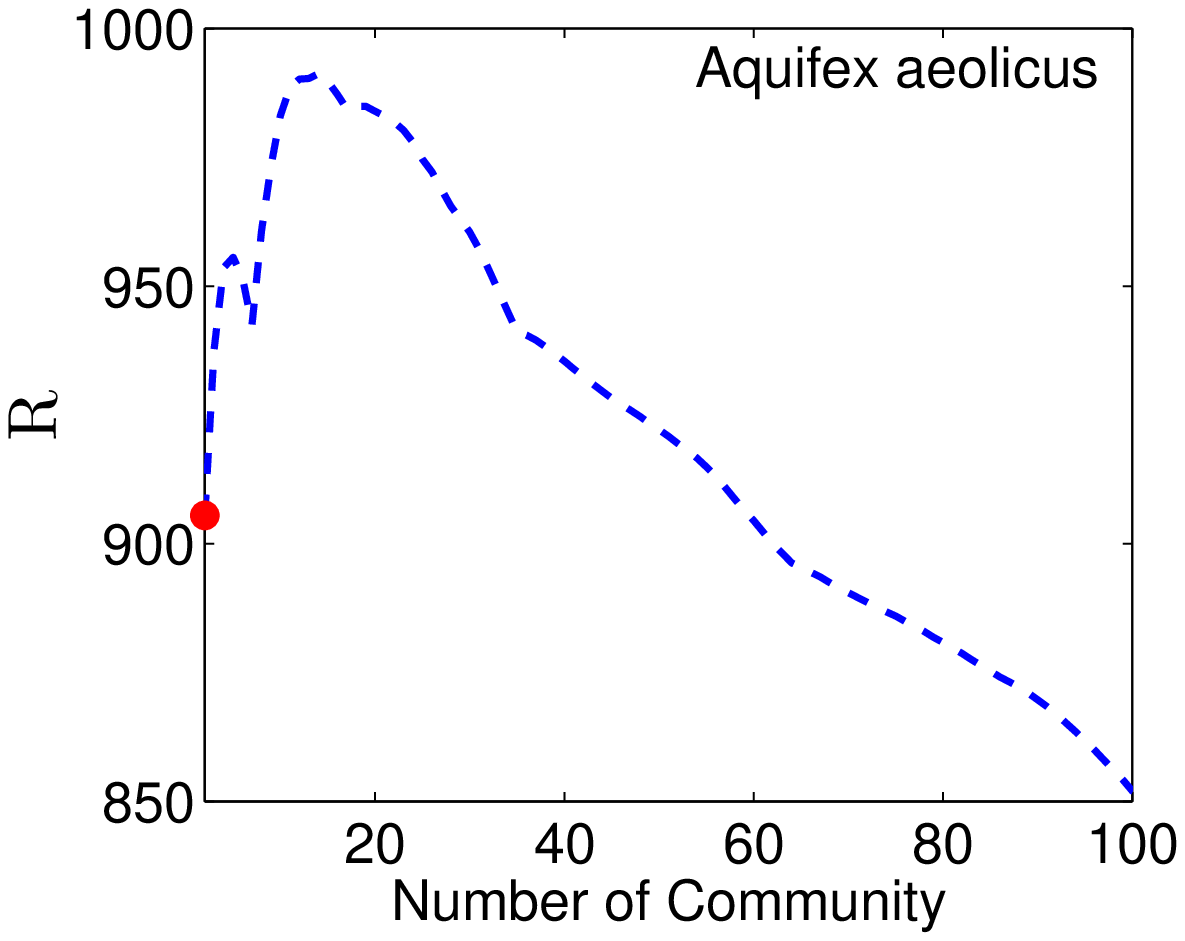}
\includegraphics[width=8cm]{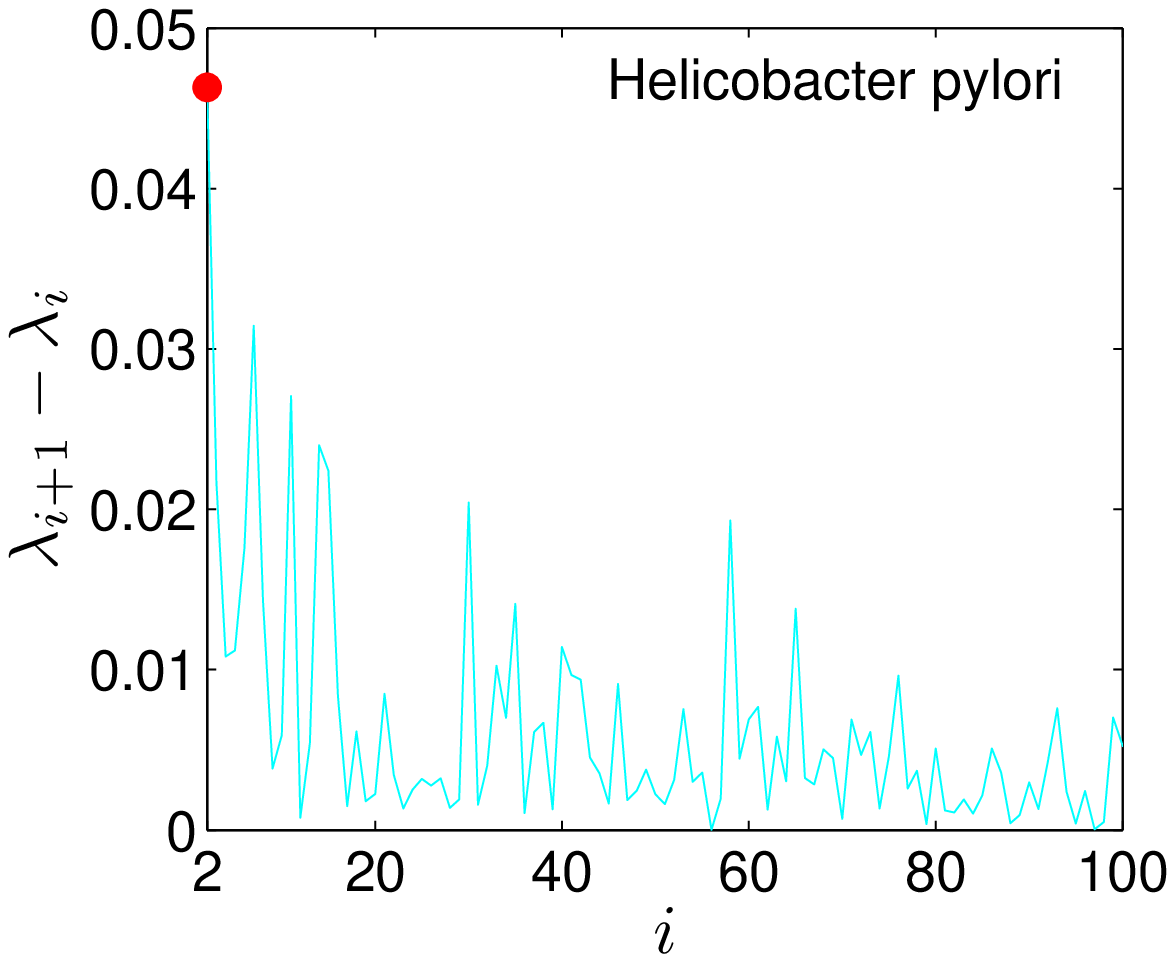}\includegraphics[width=8cm]{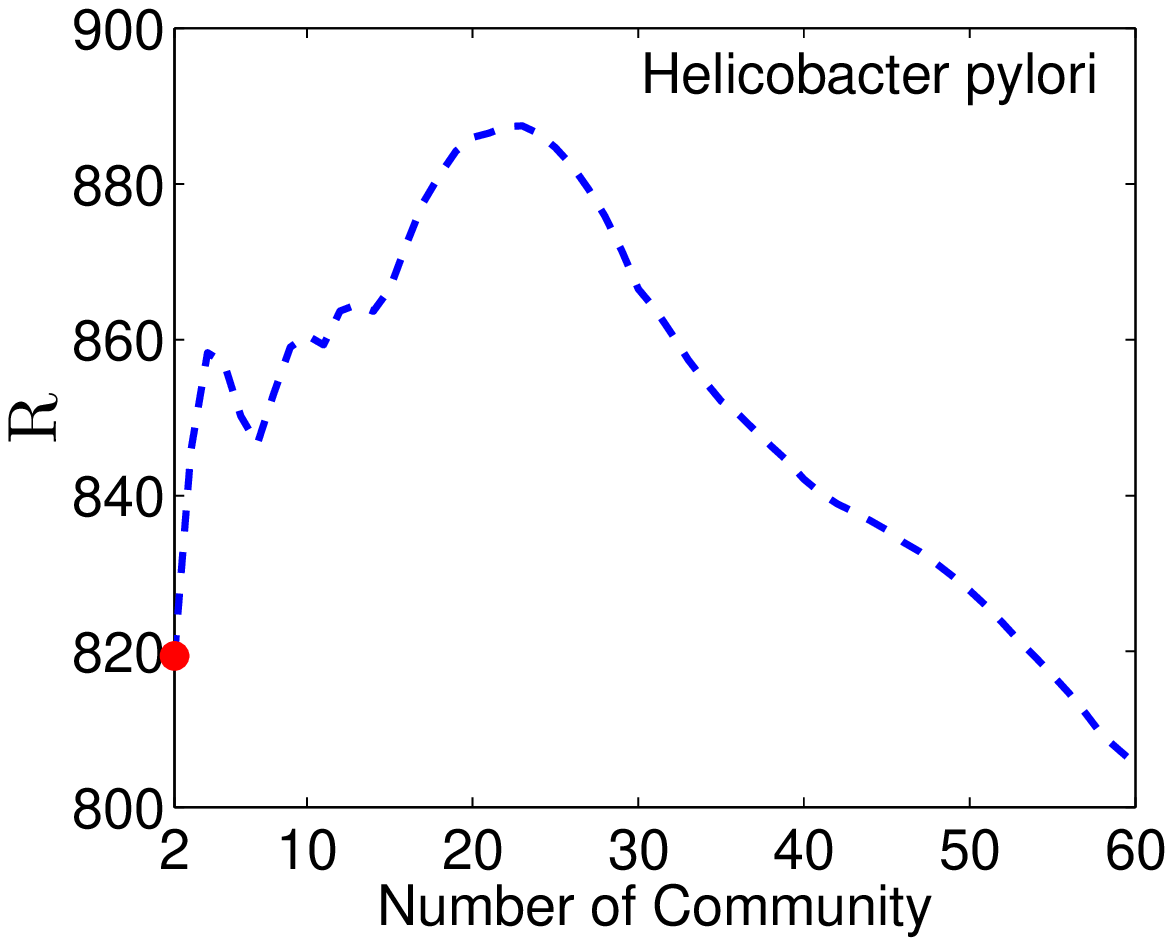}
\includegraphics[width=8cm]{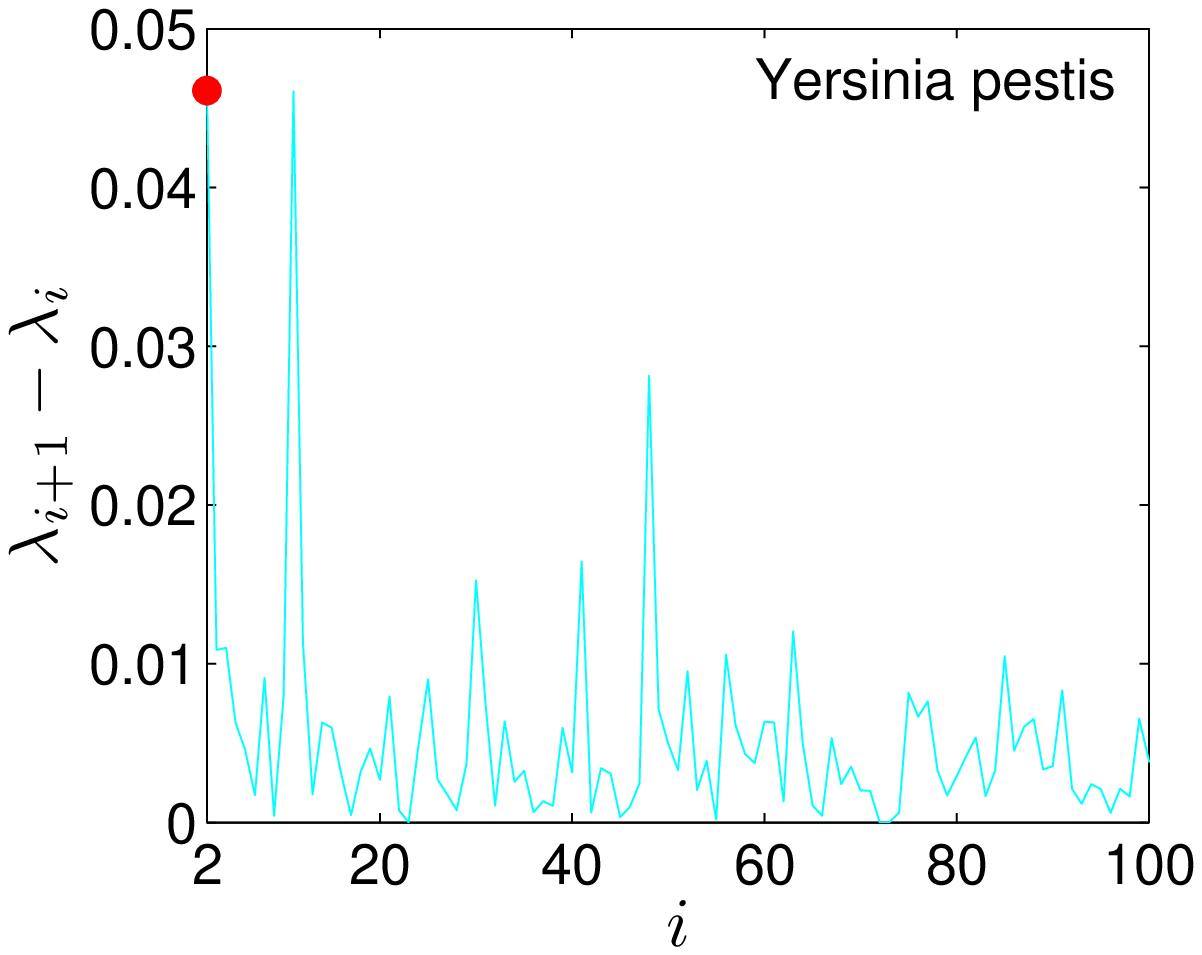}\includegraphics[width=8cm]{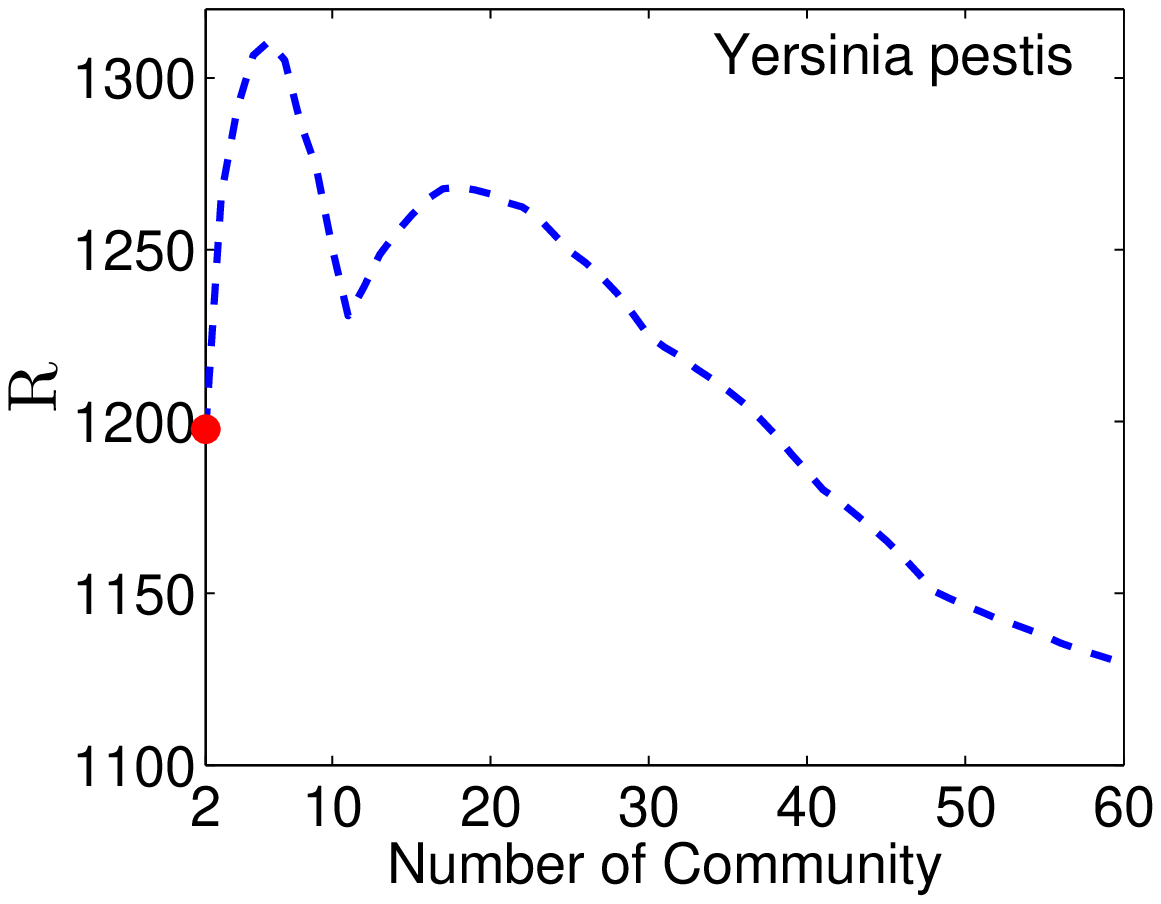}
\caption{}\label{0}
\end{figure}

\begin{figure}
\center
\includegraphics[width=8cm]{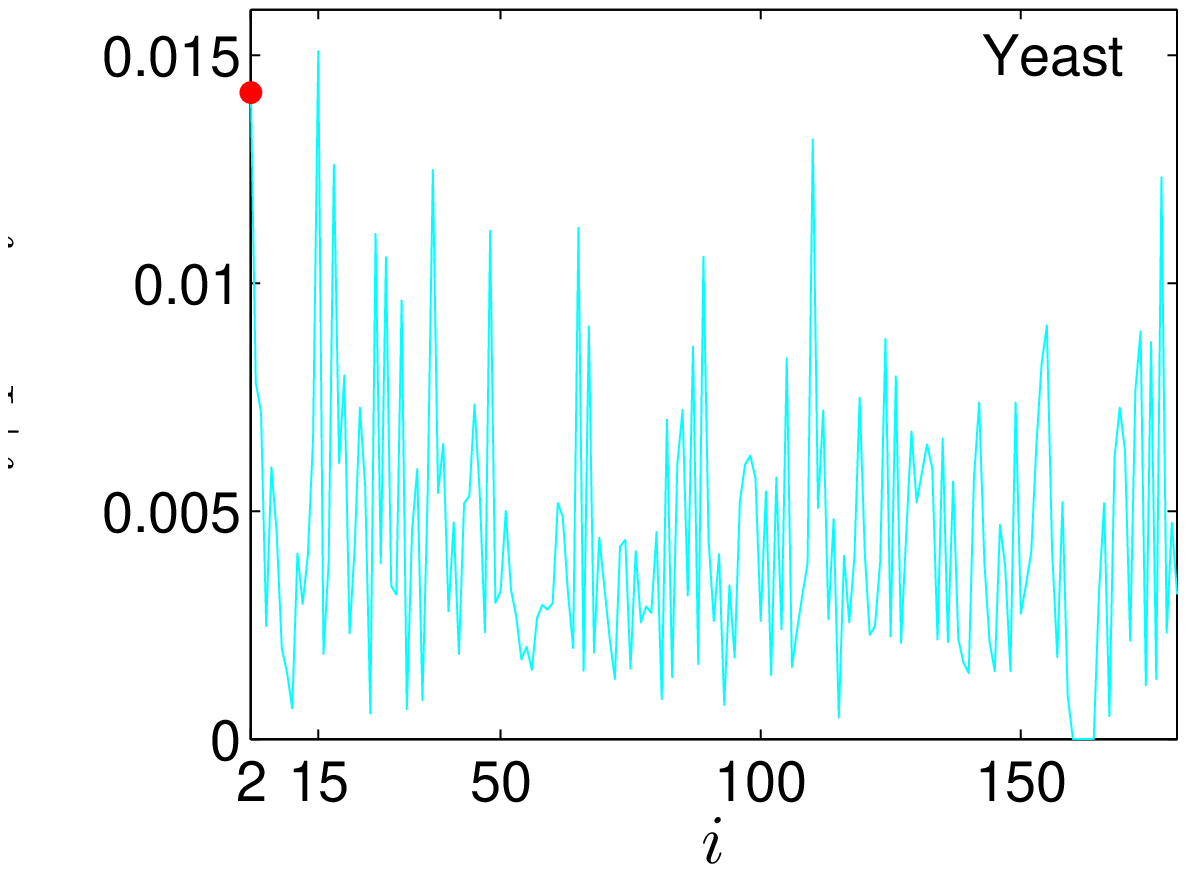}\includegraphics[width=8cm]{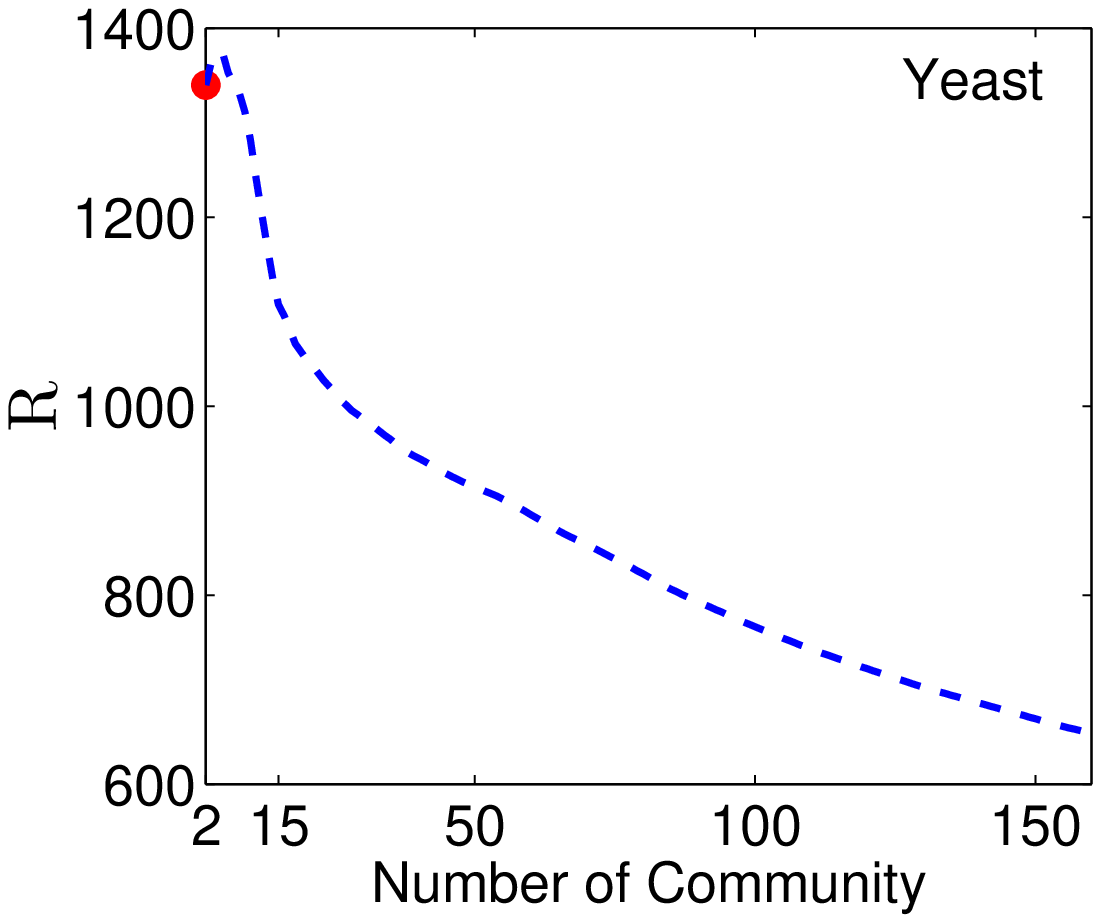}
\includegraphics[width=8cm]{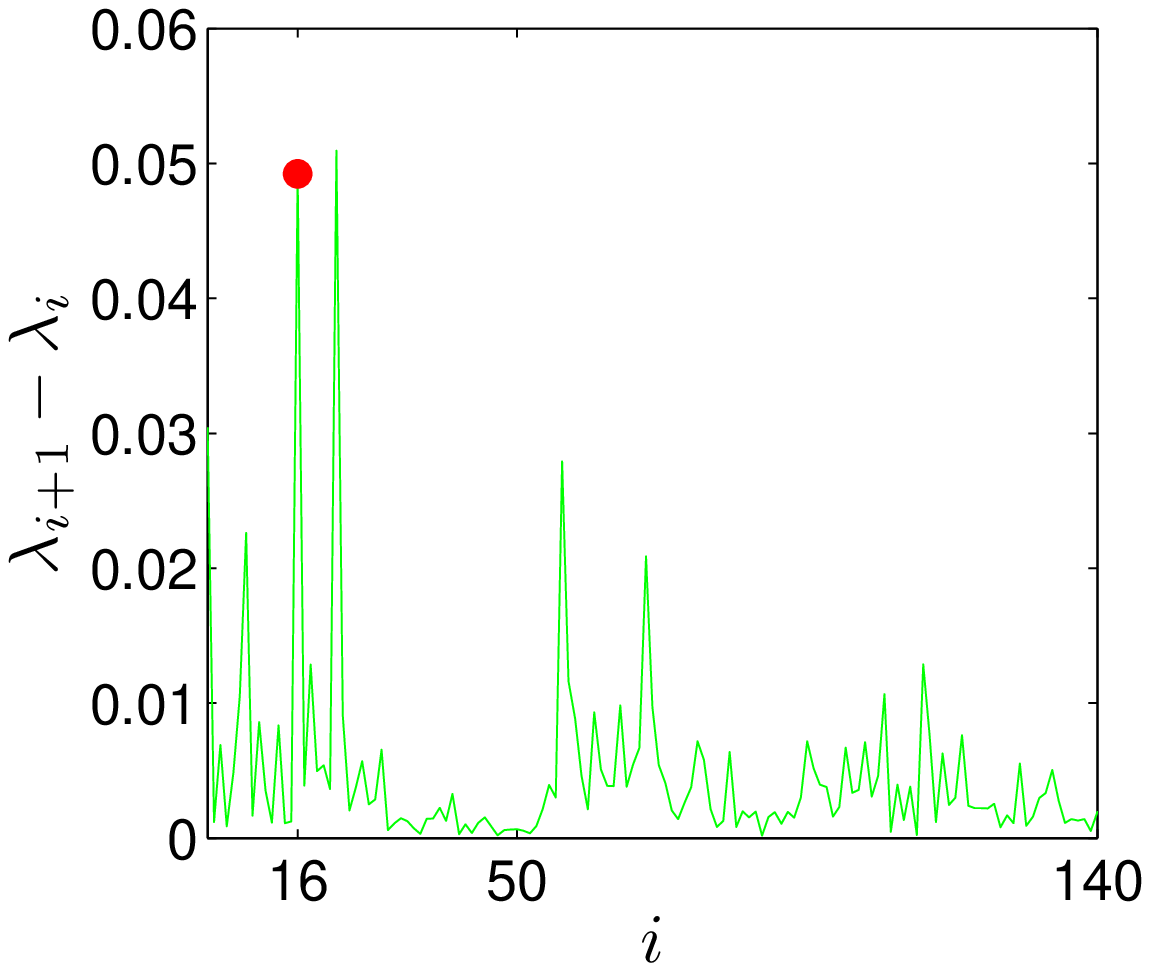}\includegraphics[width=8cm]{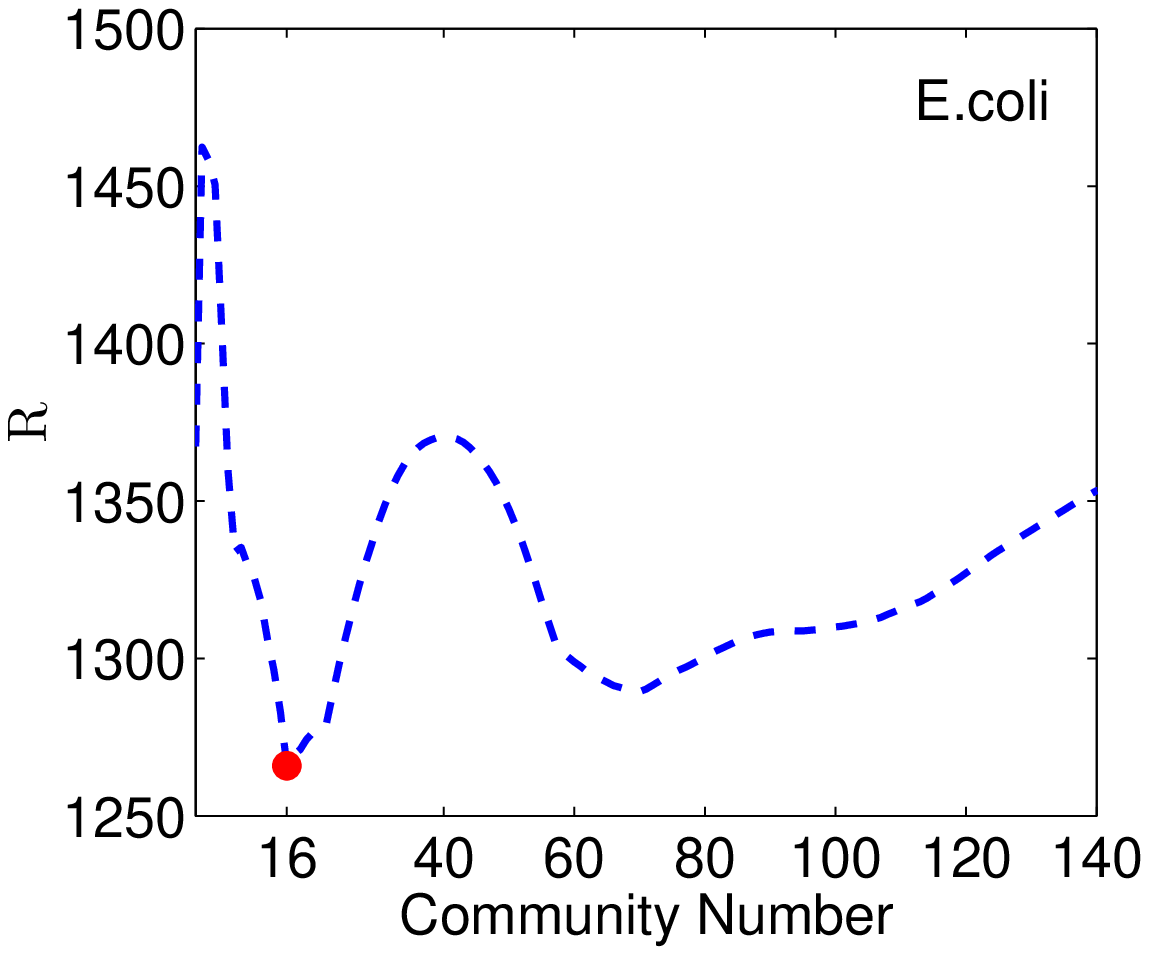}
\includegraphics[width=8cm]{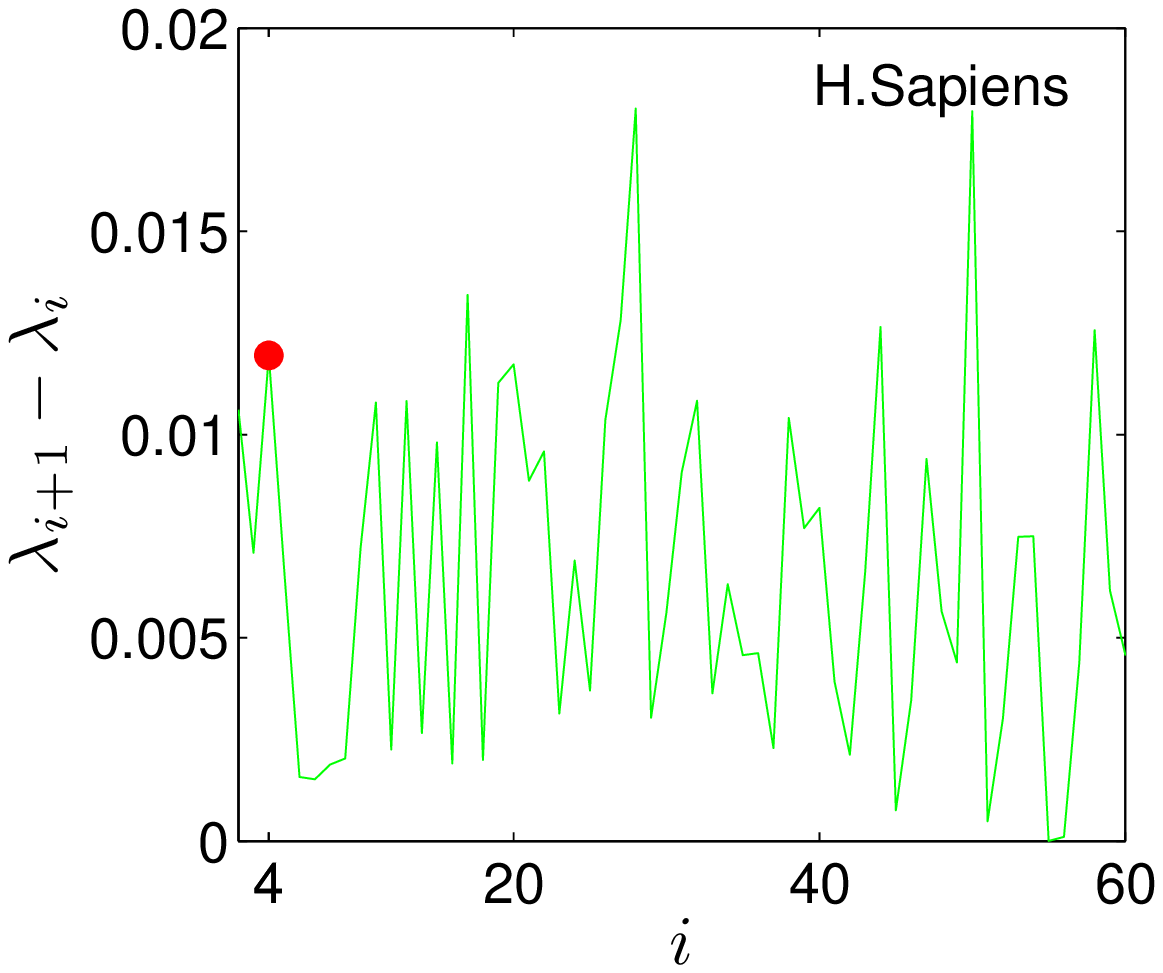}\includegraphics[width=8cm]{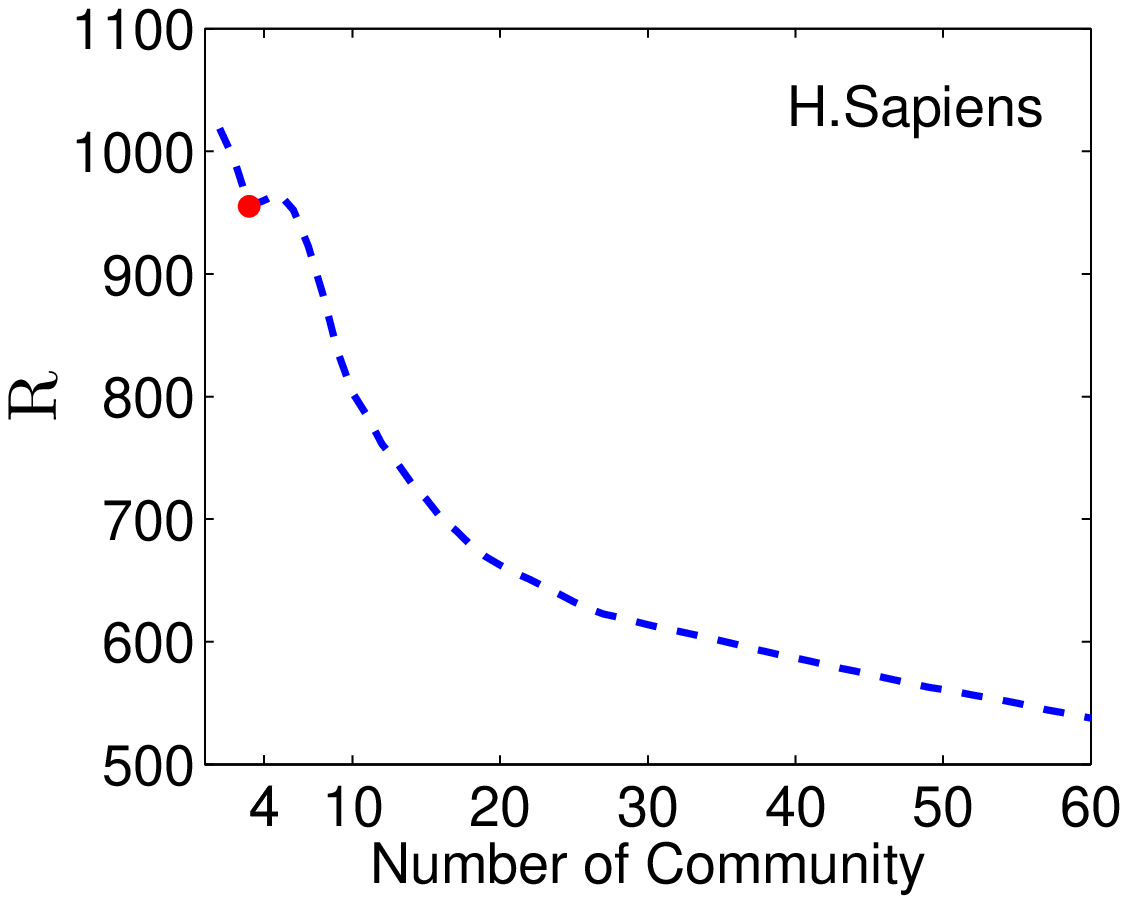}
\includegraphics[width=8cm]{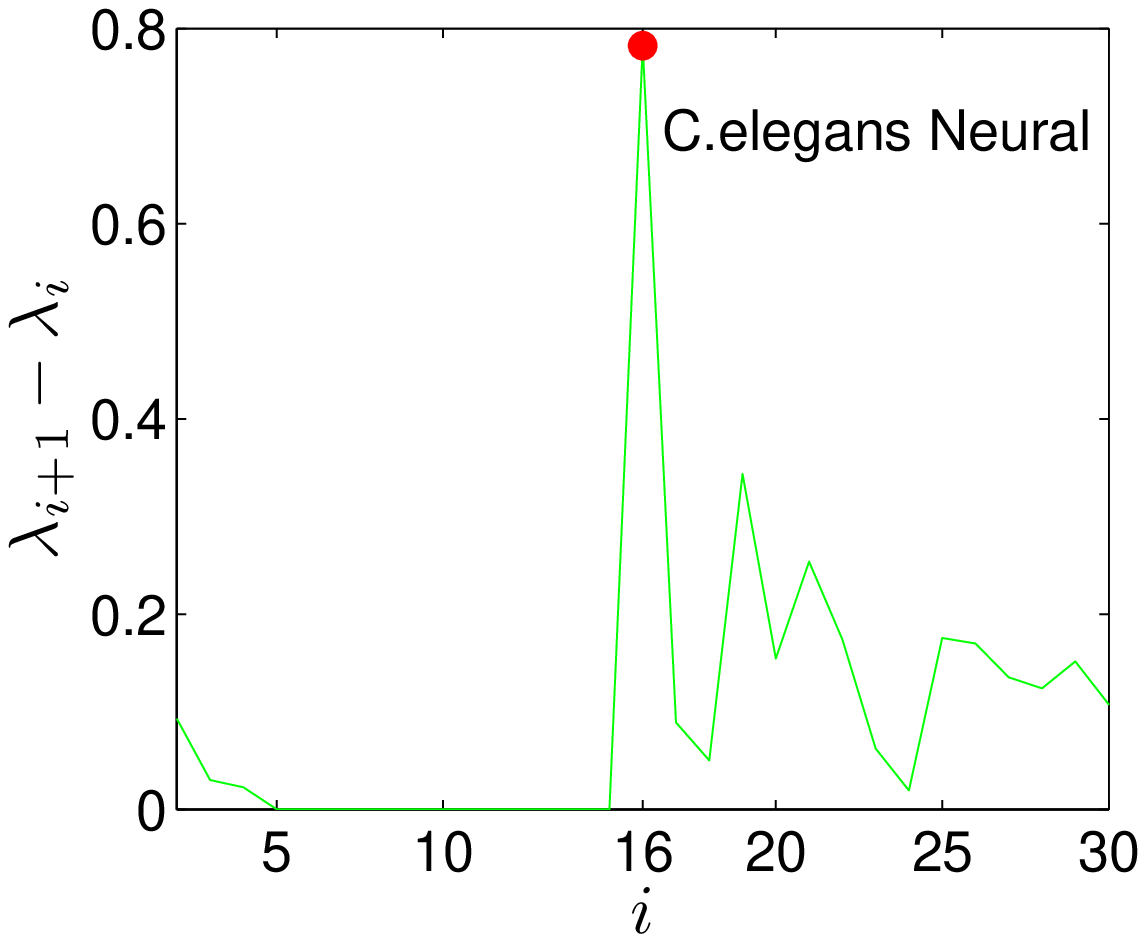}\includegraphics[width=8cm]{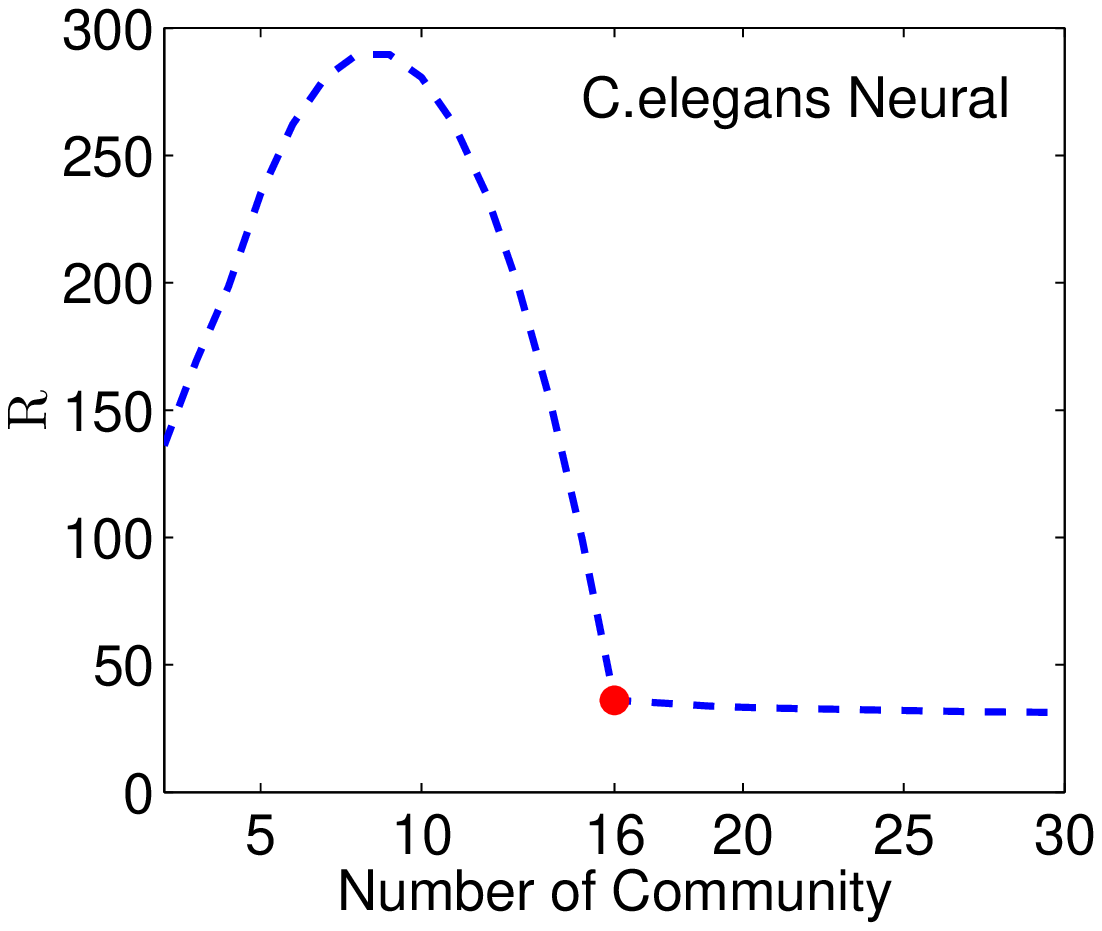}
\caption{}\label{1}
\end{figure}

\begin{figure}
\center
\includegraphics[width=8cm]{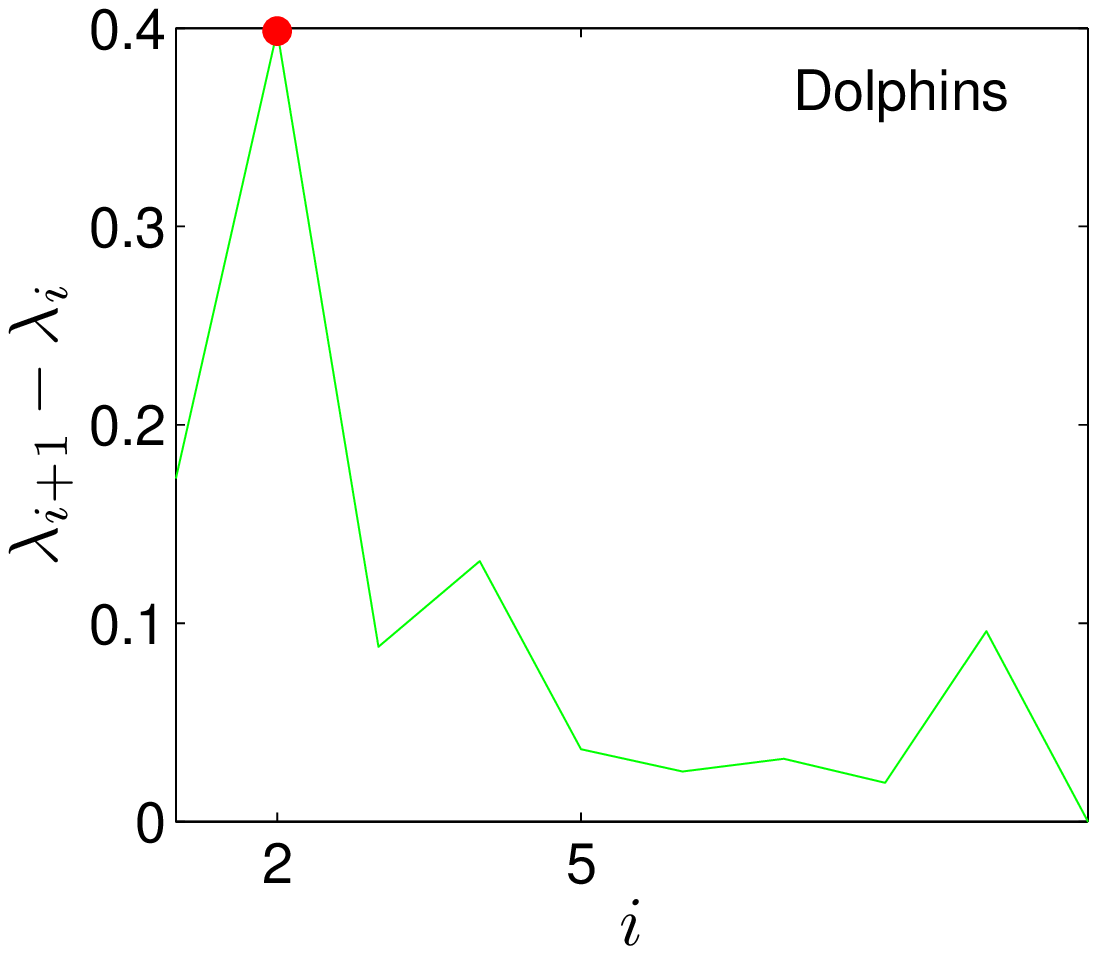 }\includegraphics[width=8cm]{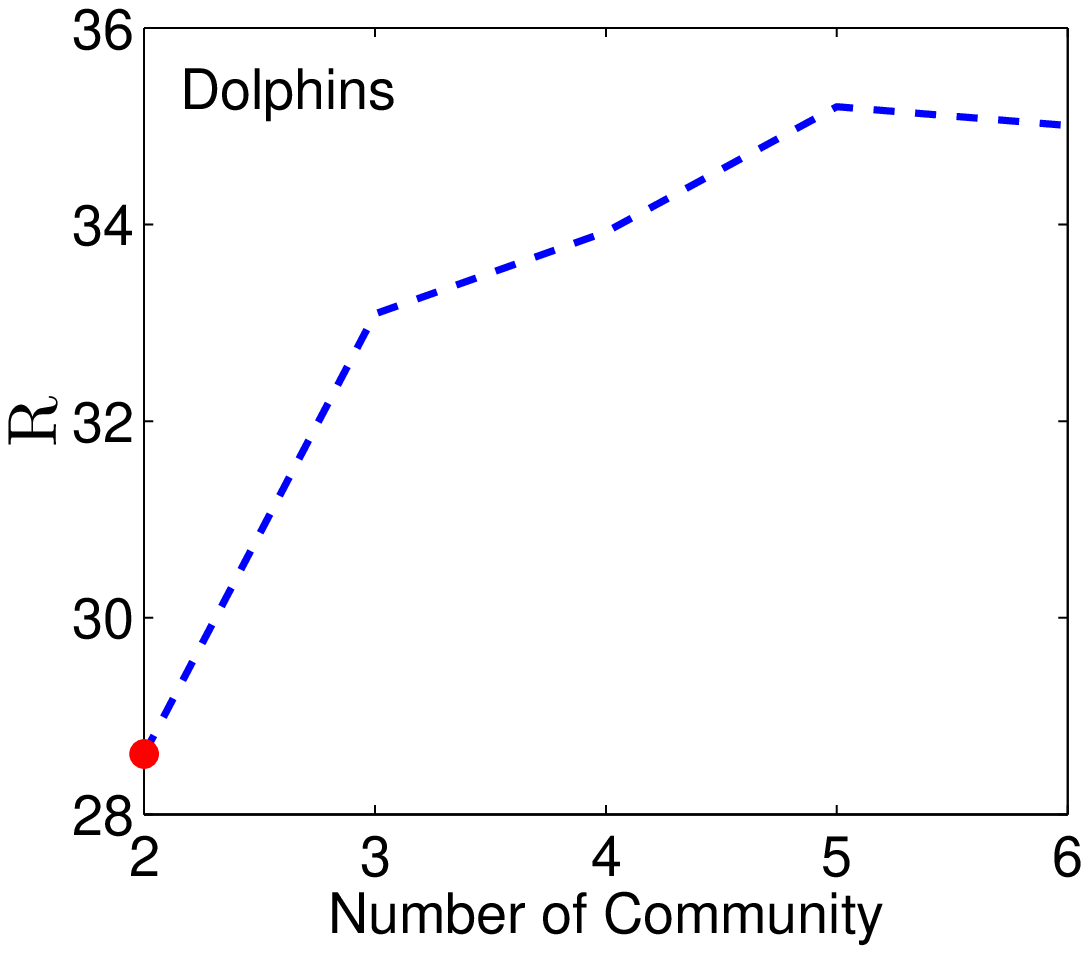}
\includegraphics[width=8cm]{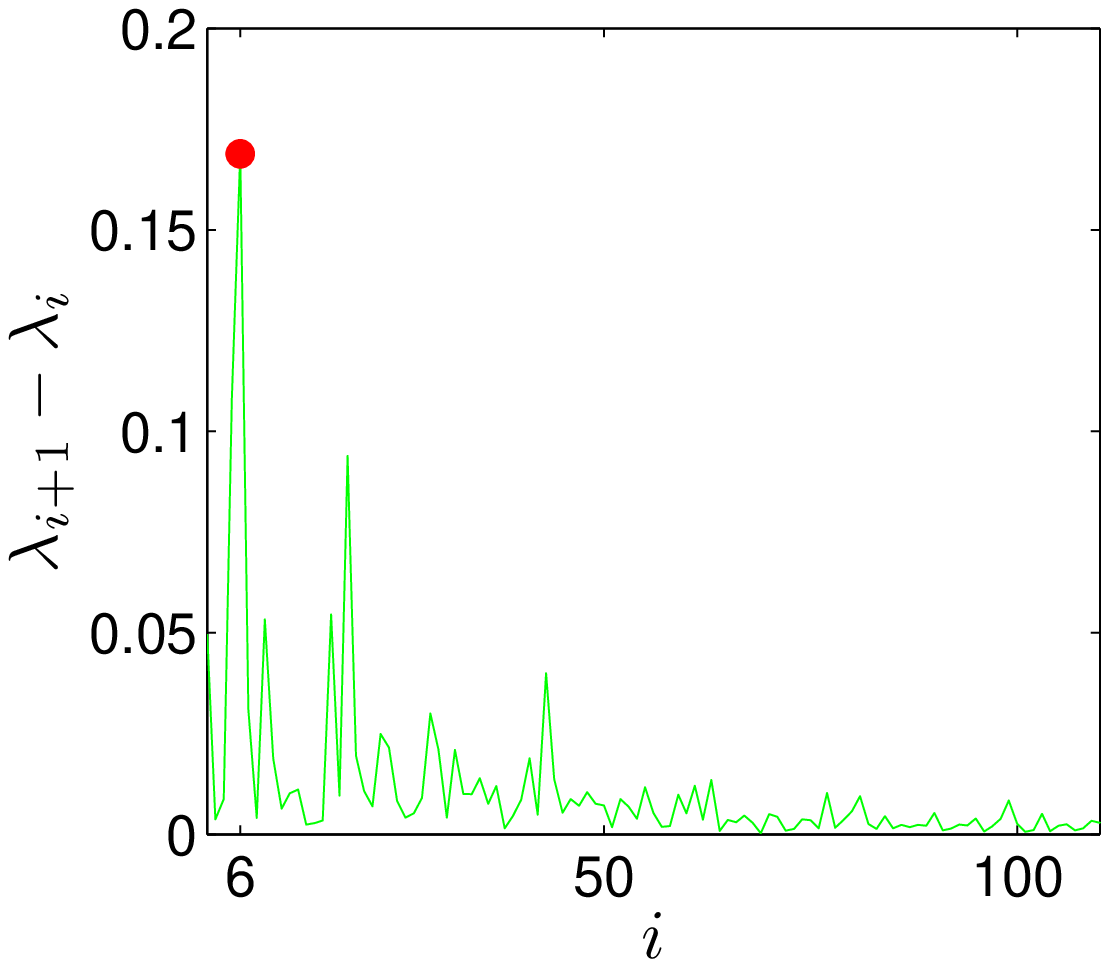}\includegraphics[width=8cm]{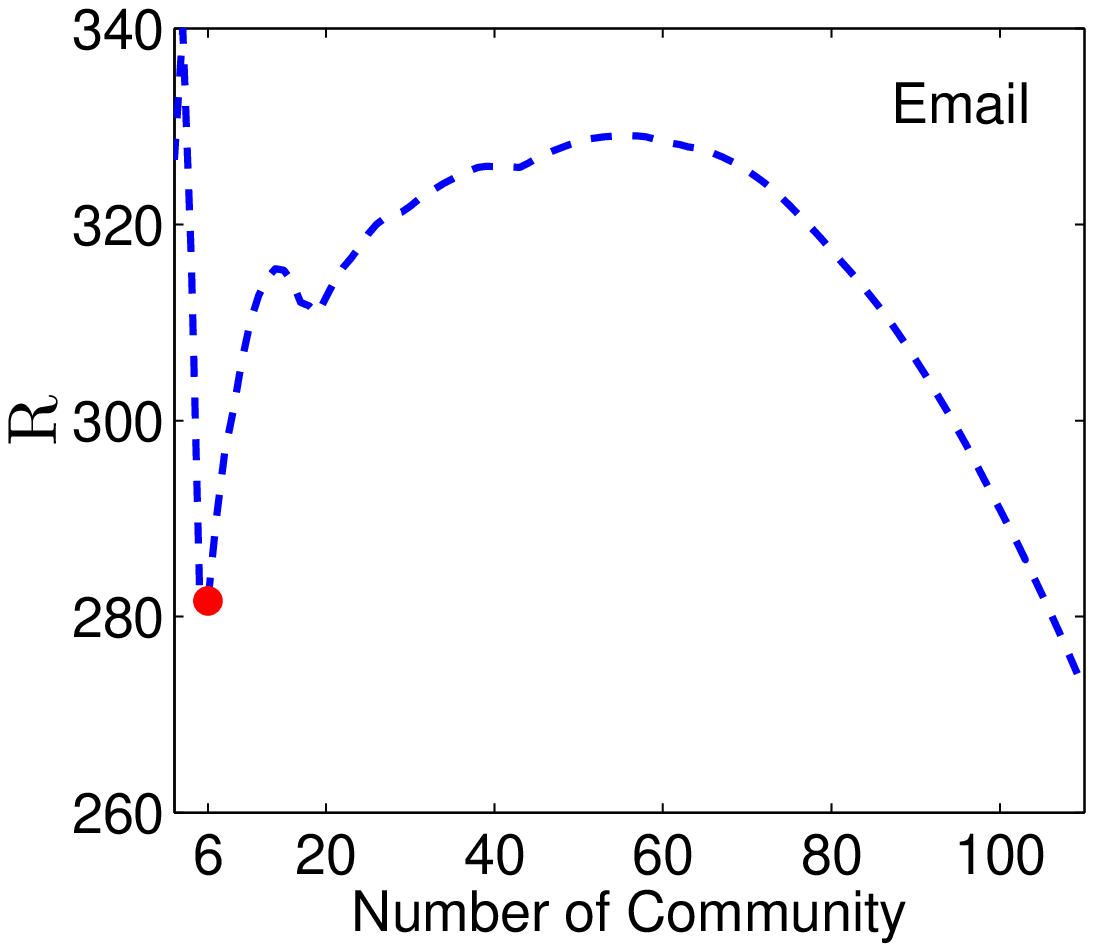}
\includegraphics[width=8cm]{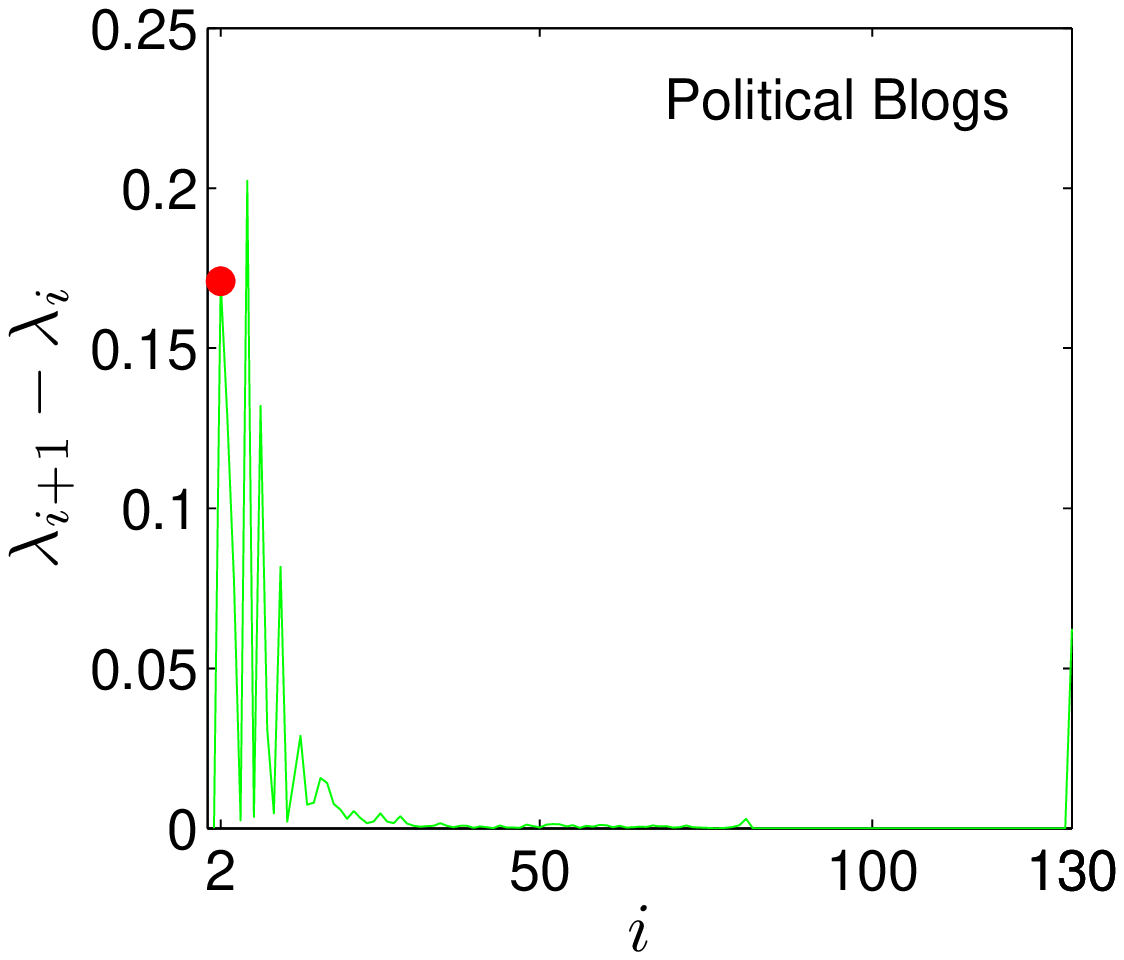}\includegraphics[width=8cm]{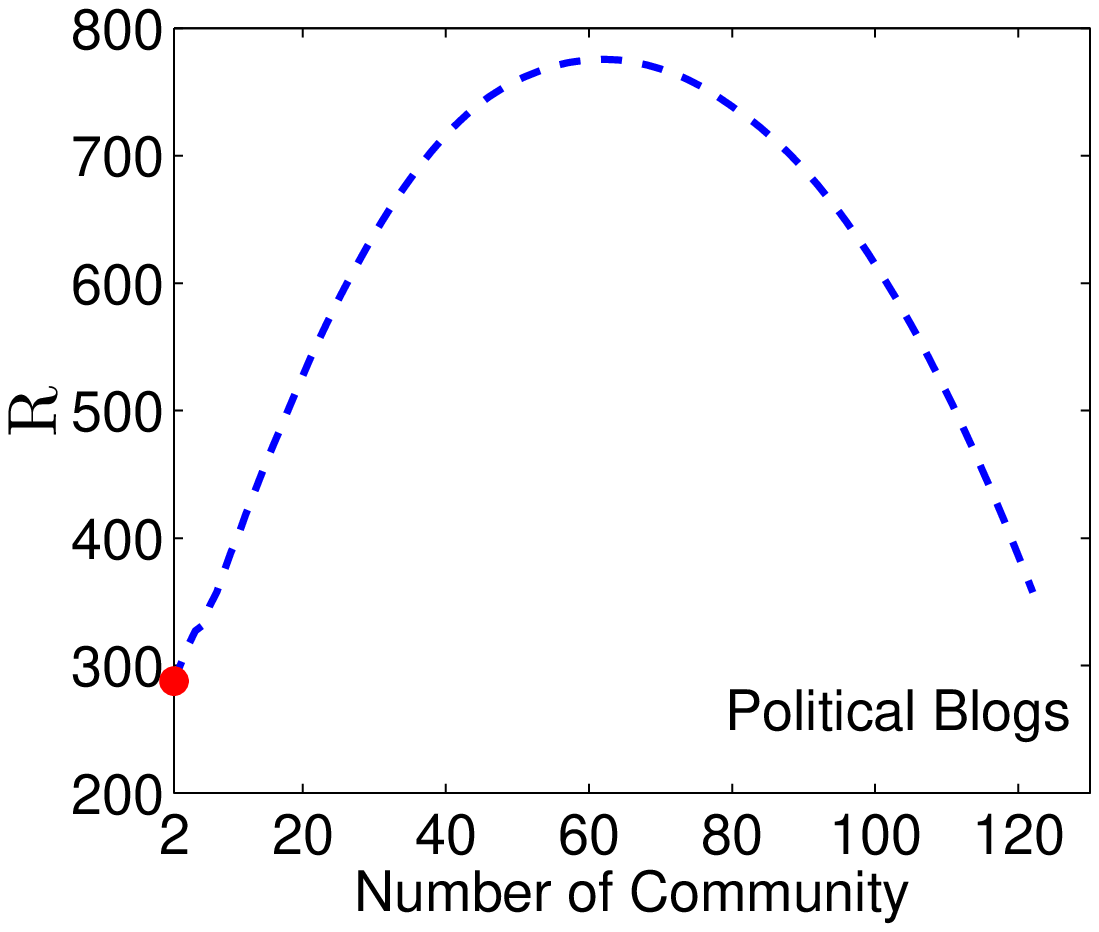}
\includegraphics[width=8cm]{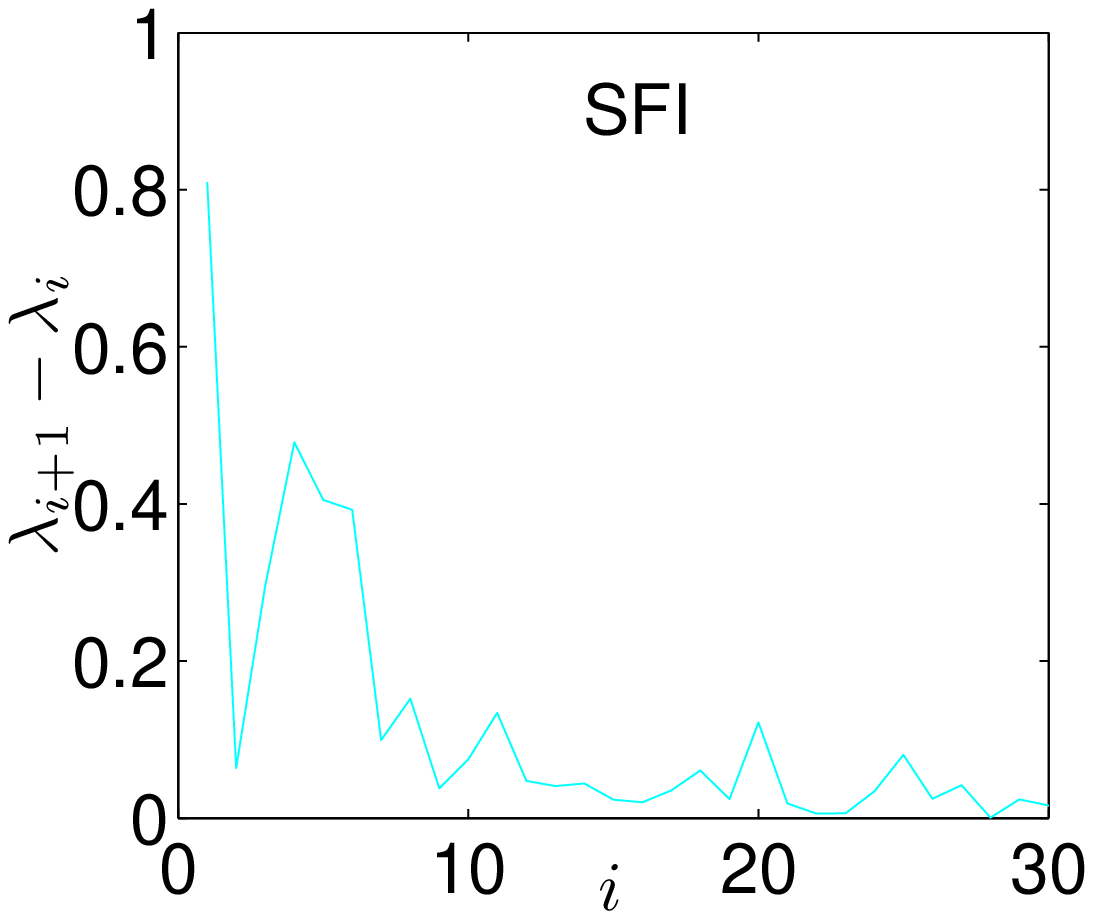}\includegraphics[width=8cm]{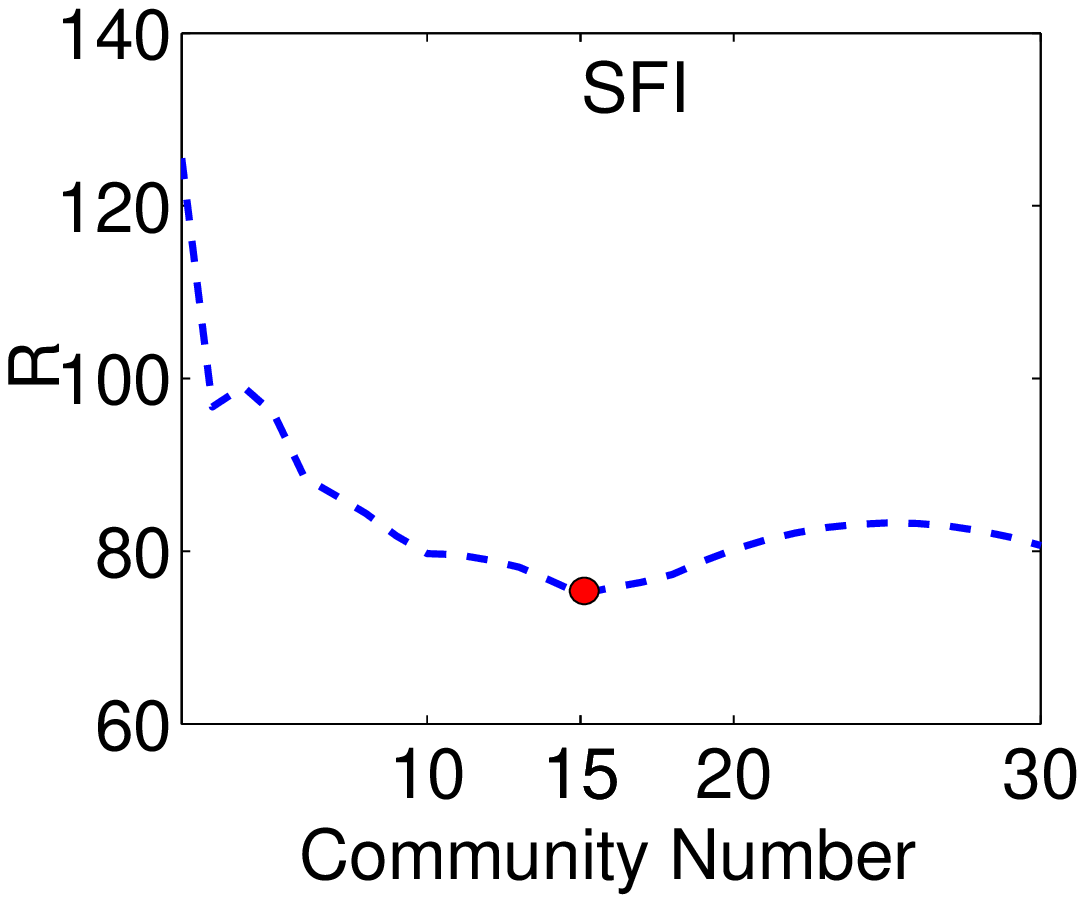}
\caption{}\label{2}
\end{figure}


\begin{thebibliography}{99}
\bibitem{SantoArxiv} S. Fortunato, arXiv:0906.0612v1, (2009).
\bibitem{linear_time}Wu, F. and Huberman, B. A. (2004) Finding communities in linear time: a physics approach. \textit{Eur. Phys. J. B}. 38:331-338.
\bibitem{newman_spectra}Newman, M. E. J. (2006) Finding community structure in networks using the eigenvectors of matrices. \textit{Phys. Rev. E}. 74: 036104.
\bibitem{GN}Girvan, M. and Newman, M. E. J. (2002) Community structure in social and biological networks. \textit{Proc. Natl. Acad}. 99: 7821-7826.
\bibitem{Systematic}Donetti, L. and  Munoz, M. A. (2004) Detecting network communities: a new systematic and efficient algorithm. \textit{J. Stat. Mech}. P10012.
\bibitem{WLC} Wang, X., Li, X. and Cheng, G. (2006) Complex network
thory and application. \textit{Tsinghua University Press.} Page
162-193.
\bibitem{Newman_Q}Newman, M. E. J. (2006) Modularity and community structure in networks. \textit{Proc. Natl. Acad.} 103: 8577-8582

\bibitem{Fan}Fan, Y., Li, M., Zhang, P., Wu, J. and Di, Z. (2007) Accuracy and precision of methods for community identification in weighted networks. \textit{Physica A}
377: 363-372 .


\bibitem{BianconiEntropy} G. Bianconi, G., Pin,  P. and Marsili, M. (2009) Assessing the relevance of node features for network
structure. \textit{Proc. Natl. Acad}. 106: 11433-11438.

\bibitem{Gfeller}Gfeller, D., Chappelier, J.-C. and de Los Rios, P. (2005) Finding instabilities in the community strucuture of complex networks. \textit{Phys. Rev. E} 72: 056135.

\bibitem{Yhu}Hu, Y., Nie, Y., Yang, Y., Cheng, J., Fan, Y. and Di,
Z. Measuring Significance of Community Structure in Complex
Networks. arXiv:0906.0493, (2009)
%\bibitem{PRE016110} J. Reichardt, S. Bornholdt, (2006) \textit{Phys. Rev. E.} 74: 016110.
\bibitem{Num_spec}McGraw, P. N. and Menzinger, M.  (2008) Laplacian spectra as a diagnostic tool for network structure and
dynamics. \textit{Phys. Rev. E.} 77: 031102.
\bibitem{Laplacian_spectra} Dorogovtsev, S. N., Goltsev, A. V., Mendes, J. F.  and
Samukhin, A. N. (2003) Spectra of complex networks. \textit{Phys.
Rev. E} 68: 046109.
%\bibitem{numerical_algbar}Jiang, C. (2003) Linear algebra calculation
%method. \textit{China Science and Technology University Press}
%Chinese Book, P 291.
\bibitem{Russia}Faddeev I, D. K., FaDDeeva, V. N. (1965) Calculation method of linear
algebra. \textit{Shanghai Science and Technology Press} (Translated
into Chinese by Li, G. \textit{et al}).
\bibitem{LFR}Lancichinetti, A., Fortunato, F. and Radicchi,
F. (2008) Benchmark graphs for testing community detection
algorithms. \textit{Phys. Rev. E.} 78: 046110.

% network data
\bibitem{ZK_d}Zachary, W. W. \textit{Journal of Anthropological Research} 33: 452-473
(1977).
\bibitem{Dolphin}Lusseau, D., Schneider, K.,  Boisseau, O. J.,  Haase, P., Slooten, E. and Dawson, S. M. (2003) The bottlenose dolphin community of Doubtful Sound features
a large proportion of long-lasting associations. \textit{Behavioral
Ecology and Sociobiology} 54: 396-405.
\bibitem{Ploblog}Adamic, L. A. and Glance, N. (2005) The Political Blogosphere and the 2004 U.S. Election:
Divided They Blog. \textit{The political blogosphere and the 2004 US
Election, in Proceedings of the WWW-2005 Workshop on the Weblogging
Ecosystem.}
\bibitem{C.neural}Watts,  D. J. and Strogatz, S. H. (1998) Collective dynamics of 'small-world'networks. \textit{Nature} 393: 440-442.
\bibitem{Jazz}Gleiser, P. and L. Danon, L. (2003) Community structure in jazz. \textit{Adv. Complex Syst.} 6: 565.
\bibitem{C.met}J. Duch and A. Arenas, \textit{Phys. Rev. E.} 72: 027104, (2005).
\bibitem{webset1}http://www-personal.umich.edu/~mejn/netdata/
\bibitem{webset2}Database of Interacting Proteins (DIP). http://dip.doe-mbi.ucla.edu
\bibitem{webset3}http://www.nd.edu/¡«networks

% network data
\end{thebibliography}
\end{document}